\documentclass[12pt]{article}
\usepackage{epsfig}
\usepackage{amssymb}
\usepackage{amsmath}
\usepackage{amsfonts}
\usepackage{graphicx}
\usepackage{mathrsfs}
\usepackage[dvips]{color}
\usepackage{multirow}

\newcommand{\bsigma}{\boldsymbol{\sigma}}

\newcommand{\R}{\mathbb{R}}
\newcommand{\C}{\mathbb{C}}

\newcommand{\fa}{\mathfrak{a}}
\newcommand{\fb}{\mathfrak{b}}
\newcommand{\fc}{\mathfrak{c}}

\newcommand{\fh}{\mathfrak{h}}

\newcommand{\fq}{\mathfrak{q}}

\newcommand{\fz}{\mathfrak{z}}

\newcommand{\bzero}{\mathbf{0}}

\newcommand{\bG}{\mathbf{G}}
\newcommand{\bH}{\mathbf{H}}
\newcommand{\bI}{\mathbf{I}}

\newcommand{\bM}{\mathbf{M}}
\newcommand{\bcM}{\boldsymbol{\cM}}

\newcommand{\bU}{\mathbf{U}}

\newcommand{\cA}{{\mathcal{A}}}
\newcommand{\cB}{\mathcal{B}}
\newcommand{\cC}{\mathcal{C}}

\newcommand{\cH}{\mathcal{H}}

\newcommand{\cF}{\mathcal{F}}
\newcommand{\cG}{\mathcal{G}}
\newcommand{\cI}{\mathcal{I}}

\newcommand{\cK}{\mathcal{K}}

\newcommand{\cM}{\mathcal{M}}

\newcommand{\cR}{\mathcal{R}}

\newcommand{\cT}{\mathcal{T}}
\newcommand{\cU}{\mathcal{U}}

\newcommand{\be}{\begin{equation}}
\newcommand{\ee}{\end{equation}}
\newcommand{\bea}{\begin{eqnarray}}
\newcommand{\eea}{\end{eqnarray}}
\newcommand{\nn}{\nonumber}

\newcommand{\ed}{\end{document}}

\newcommand{\bi}{\begin{itemize}}
\newcommand{\ei}{\end{itemize}}

\newcommand{\bce}{\begin{center}}
\newcommand{\ece}{\end{center}}

\newcommand{\sH}{\mathscr{H}}

\newcommand{\sT}{\mathscr{T}}
\newcommand{\sU}{\mathscr{U}}
\newcommand{\RE}{{\rm Re}}
\newcommand{\IM}{{\rm Im}}

\newcommand{\bcK}{{\boldsymbol{\cK}}}

\newcommand{\bcF}{{\boldsymbol{\cF}}}
\newcommand{\bcG}{{\boldsymbol{\cG}}}
\newcommand{\bcH}{{\boldsymbol{\cH}}}

\newcommand{\bcU}{{\boldsymbol{\cU}}}

\newcommand{\brv}{\breve}

\begin{document}

\title{Transfer matrix for long-range potentials}

\author{Farhang~Loran\thanks{E-mail address: loran@iut.ac.ir}~
and Ali~Mostafazadeh\thanks{E-mail address:
amostafazadeh@ku.edu.tr}\\[6pt]
$^{*}$Department of Physics, Isfahan University of Technology, \\ Isfahan 84156-83111, Iran\\[6pt]
$^\dagger$Departments of Mathematics and Physics, Ko\c{c} University,\\  34450 Sar{\i}yer,
Istanbul, Turkey}
\date{ }
\maketitle

\begin{abstract}
We extend the notion of the transfer matrix of potential scattering
to a large class of long-range potentials $v(x)$ and derive its
basic properties. We outline a dynamical formulation of the
time-independent scattering theory for this class of potentials
where we identify their transfer matrix with the $S$-matrix of a
certain effective non-unitary two-level quantum system. For
sufficiently large values of $|x|$, we express $v(x)$ as the sum of
a short-range potential and an exactly solvable long-range
potential. Using this result and the composition property of the
transfer matrix, we outline an approximation scheme for solving the
scattering problem for $v(x)$. To demonstrate the effectiveness of
this scheme, we construct an exactly solvable long-range potential
and compare the exact values of its reflection and transmission
coefficients with those we obtain using our approximation scheme.




\end{abstract}

\section{Introduction}

Transfer matrices have been introduced and employed as a powerful tool for dealing with scattering problems for finite-range locally periodic potentials \cite{jones-1941,abeles,thompson}. These typically arise in the study of wave propagation in multilayered media \cite{yeh,pereyra,griffiths,yeh-book}. By definition, a function $v:\R\to\C$ is called a short-range potential \cite{yafaev}, if there are positive real numbers $C$, $D$, and $\alpha$ such that $\alpha>1$ and
    \be
    |v(x)|\leq\frac{C}{\left(1+|x|\right)^\alpha}~~~{\rm for}~~~|x|\geq D.
    \label{condi-0}
    \ee
An important consequence of this condition is that, as $x\to\pm\infty$, the solutions of the time-independent Schr\"odinger equation,
    \be
    -\psi''(x)+v(x)\psi(x)=k^2\psi(x),
    \label{sch-eq}
    \ee
tend to plane waves, i.e.,
    \be
    \psi(x)\to A_\pm e^{ikx}+B_\pm e^{-ikx}~~{\rm for}~~x\to\pm\infty.
    \label{e2-SR}
    \ee
This in turn allows for the introduction of the transfer matrix of $v$ as a $2\times 2$ matrix $\bM$ that satisfies
     \be
    \left[\begin{array}{cc}
    A_+\\B_+\end{array}\right]
    =\bM\left[\begin{array}{cc}
    A_-\\B_-\end{array}\right].
    \label{M-def}
    \ee
This condition determines $\bM$ in a unique manner provided that we demand that it is independent of $A_-$ and $B_-$ \cite{epjp-2019}.

The transfer matrix has two important properties \cite{sanchez,bookchapter}:
    \begin{enumerate}
    \item Its entries, $M_{ij}$, determine the left/right reflection ($R^{l/r}$) and transmission ($T^{l/r}$) amplitudes according to
    \begin{align}
    &R^l=-\frac{M_{21}}{M_{22}},
    &&R^r=\frac{M_{12}}{M_{22}},
    &&T^l=T^r=T:=\frac{1}{M_{22}}.
    \label{RT}
    \end{align}
    \item For any strictly increasing finite sequence of real numbers, $a_1,a_2,a_3,\cdots,a_{n-1}$, and potentials $v_j$ defined by
        \be
        v_j(x):=\left\{\begin{array}{cc}
        v(x) & {\rm for}~x\in (a_{j-1},a_{j}),\\[6pt]
        0 & {\rm otherwise},\end{array}\right.
        \label{vj-def}
        \ee
with $j\in\{1,2,\cdots,n\}$, $a_0:=-\infty$, and $a_n:=+\infty$, we can express $\bM$ in terms of the transfer matrices $\bM_j$ of $v_j$ according to
    \be
    \bM=\bM_n\bM_{n-1}\cdots\bM_1.
    \label{composition}
    \ee
    \end{enumerate}

Equation~(\ref{composition}), which is known as the composition
property of the transfer matrix, allows for the reduction of the
solution of the scattering problem for a given scatterer to that of
its slices along the scattering axis. This is the main reason for
the practical significance of the transfer matrix \cite{griffiths}
and its generalizations to multichannel
\cite{pereyray-1998a,pereyray-2002,pereyray-2005,Shukla-2005,anzaldo-meneses-2007},
multidimensional \cite{pendry-1984,pendry-1990a,pra-2016}, and
electromagnetic
\cite{teitler-1970,berreman-1972,pendry-1994,mclean,ward-1996,pendry-1996,p141}
scattering problems.

A recent observation regarding the possibility of reducing scattering problems defined on the half-line to those defined on the whole line \cite{ap-2019b} extends the domain of application of the transfer matrix (\ref{M-def}) to potentials defined on the half-line. The principle example of the latter is the effective potentials one encounters in solving the Schr\"odinger and Helmholtz equations for a spherically symmetric potential in three dimensions, i.e.,
    \be
    v_{\rm eff}(r):= \frac{l(l+1)}{r^2}+v_{\rm int}(r),
    \label{v-eff}
    \ee
where $r$ is the radial spherical coordinate, $l$ is the angular momentum quantum number, and $v_{\rm int}$ is the interaction potential. 
If $v_{\rm int}$ is a finite-range piecewise constant potential, one can express the solution of the corresponding Schr\"odinger equation in the intervals where $v_{\rm int}(r)$ is constant in terms of spherical Bessel and Hankel functions. By letting these play the role of the plane waves, $e^{\pm ikx}$, in the above discussion of the transfer matrix, one can introduce a transfer matrix capable of dealing with the scattering problem for (\ref{v-eff}), \cite{burlak}. See also \cite{moroz-2005}.

The present investigation aims at providing a systematic generalization of the notion of transfer matrix to the class $\cC_{\alpha>1/2}$ of long-range potentials that satisfy (\ref{condi-0}) for some $\alpha>1/2$. It is motivated by the above-mentioned developments related to finite-range piecewise constant spherically symmetric potentials as well as the recent discovery of long-range potentials supporting full-band unidirectional invisibility \cite{horsley,longhi-2015,hl-review,longhi-2016}.

The organization of the paper is as follows. In Sec.~\ref{S2} we reexamine the transfer matrix for a short-range potential, explore its relationship with the classical notion of the fundamental matrix of the theory of ordinary differential equations, and introduce its decomposition into a pair of matrices that respectively carry the information about the scattering properties of the potential for left- and right-incident waves. In Sec.~\ref{S3}, we give the definition of the transfer matrix for the real long-range potentials $v$ belonging to $\cC_{\alpha>1/2}$, derive its basic properties, and examine its generalization to complex long-range potentials. In Sec.~\ref{S4}, we introduce a decomposition of $v$ into the sum of a short-range potential and an exactly solvable long-range potential. This forms the basis of an approximation scheme for the solution of the scattering problem for $v$ which we outline in Sec.~\ref{S5}. In Sec.~\ref{S6}, we construct an exactly solvable long-range potential and compare the exact expression for its reflection and transmission coefficients with the outcome of our approximation scheme. Finally, in Sec.~\ref{S7}, we present our concluding remarks.

\section{Transfer and fundamental matrices for short-range potentials}
\label{S2}

Because the Schr\"odinger equation (\ref{sch-eq}) is a second-order linear homogeneous equation, its general solution $\psi$ (for each $k$) is a linear combination of a pair of linearly independent solutions,  $\psi_1$ and $\psi_2$;
    \be
    \psi(x)=c_1\psi_1(x)+c_2\psi_2(x),
    \label{gen-sol-1}
    \ee
where $c_1$ and $c_2$ are constant coefficients. Introducing the fundamental matrix \cite{ODE},
    \be
    \bcF(x):=\left[\begin{array}{cc}
    \psi_1(x)&\psi_2(x)\\
    \psi_1'(x)&\psi'_2(x)\end{array}\right],
    \label{FM-def}
    \ee
we can use (\ref{gen-sol-1}) to show that
    \be
    \left[\begin{array}{c}
    \psi(x)\\
    \psi'(x)\end{array}\right]=\bcF(x)\left[\begin{array}{c}
    c_1\\
    c_2\end{array}\right].
    \nn
    \ee
This in turn implies that for any pair of real numbers $x_\pm$,
    \be
    \left[\begin{array}{c}
    \psi(x_+)\\
    \psi'(x_+)\end{array}\right]=
    \bcF(x_+)\bcF(x_-)^{-1}
    \left[\begin{array}{c}
    \psi(x_-)\\
    \psi'(x_-)\end{array}\right].
    \label{FM-2}
    \ee

Next, we consider the case where $v$ is a short-range potential and examine the consequences of making $x_\pm$ approach $\pm\infty$. In this case, $\psi(x_\pm)$ tends to $A_\pm e^{ikx}+B_\pm e^{-ikx}$. Therefore,
    \be
    \left[\begin{array}{c}
    \psi(x_\pm)\\
    \psi'(x_\pm)\end{array}\right]\to\bcF_0(x_\pm)
    \left[\begin{array}{c}
    A_\pm\\
    B_\pm\end{array}\right]~~{\rm for}~~x_\pm\to\pm\infty,
    \label{FM-3}
    \ee
where
    \be
    \bcF_0(x):=\left[\begin{array}{cc}
    e^{ikx}&e^{-ikx}\\
    ik e^{ikx}&-ik e^{-ikx}\end{array}\right]=\bcF_0(0)e^{ikx\bsigma_3},
    \label{FM-0}
    \ee
and $\bsigma_i$ stands for the $i$-th Pauli matrix. We can use (\ref{FM-2}) and (\ref{FM-3}) to relate $\left[\begin{array}{c}
    A_-\\
    B_-\end{array}\right]$ to $\left[\begin{array}{c}
    A_+\\
    B_+\end{array}\right]$.
This reproduces (\ref{M-def}) with the transfer matrix given by
    \be
    \bM=\bcG(+\infty)\bcG(-\infty)^{-1},
    \label{M-GG}
    \ee
where
    \bea
    &&\bcG(\pm\infty):=\lim_{x\to\pm\infty}\bcG(x),
    \label{G-lim}\\
    &&\bcG(x):=\bcF_0(x)^{-1}\bcF(x).
    \label{G-def}
    \eea
Substituting (\ref{FM-def}) and (\ref{FM-0}) in (\ref{G-def}), we can identify the $j$-th column of $\bcG(x)$ with the two-component wave function:
    \bea
    \Psi_j(x)&:=&\bcF_0(x)^{-1}\left[\begin{array}{c}
    \psi_j(x)\\[6pt]
    \psi'_j(x)\end{array}\right]=
    \frac{1}{2}\left[\begin{array}{c}
    e^{-ikx}\{\psi_j(x)- i k^{-1}\psi'_j(x)\} \\[6pt]
    e^{ikx}\{\psi_j(x)+ i k^{-1}\psi'_j(x)\} \end{array}\right],
    \label{2-comp-FM}
    \eea
i.e.,
    \be
    \bcG(x)=\left[\begin{array}{cc}
    \!\Psi_1(x) & \Psi_2(x)\!\end{array}\right].
    \label{G=}
    \ee
An interesting property of $\bcG(x)$ is that its determinant is proportional to the Wronskian of the solutions $\psi_1$ and $\psi_2$;
    \be
    \det\bcG(x)=\frac{i}{2k}[\psi_1(x)\psi_2'(x)-\psi_2(x)\psi_1'(x)].
    \label{det-G}
    \ee
By virtue of Abel's theorem \cite{ODE}, this implies that $\det \bcG(x)$ does not depend on $x$. Furthermore, because $\psi_1$ and $\psi_2$ are linearly independent solutions of the Schr\"odinger equation, their Wronskian and consequently $\det \bcG(x)$ do not vanish \cite{ODE}.

The analysis leading to the decomposition (\ref{M-GG}) of the transfer matrix is valid for every linearly independent pair of solutions, $\psi_1$ and $\psi_2$, of the Schr\"odinger equation~(\ref{sch-eq}). If we identify these respectively with the left- and right-incident scattering solutions, $\psi_l$ and $\psi_r$, which by definition satisfy
    \bea
    \psi_l(x)&\to&\left\{\begin{array}{ccc}
    e^{ikx} + R^l e^{-ikx} &{\rm for}&x\to-\infty,\\
    T^l e^{ikx} &{\rm for}&x\to+\infty,\end{array}\right.
    \label{psi-left-FM}\\[3pt]
    \psi_r(x)&\to&\left\{\begin{array}{ccc}
    T^r e^{-ikx} &{\rm for}&x\to-\infty,\\
    e^{-ikx} + R^r e^{ikx}&{\rm for}&x\to+\infty,\end{array}\right.
    \label{psi-right-FM}
    \eea
and make use of (\ref{G-lim}) -- (\ref{G=}) to compute $\bcG(\pm\infty)$, we find
    \begin{align}
    &\bcG(-\infty)=\left[\begin{array}{cc}
    1 & 0\\
    R^l & T^r\end{array}\right],
    &&
    \bcG(+\infty)=\left[\begin{array}{cc}
    T^l & R^r\\
    0 & 1\end{array}\right].
    \label{GG=}
    \end{align}
Because $\bcG(x)$ has a constant nonzero determinant, these equations imply the transmission reciprocity,
    \be
    T^r=T^l,
    \label{reciprocity}
    \ee
and the impossibility of perfect absorption \cite{bookchapter},
    \be
    T_\pm\neq 0.
    \ee
According to (\ref{GG=}) and (\ref{reciprocity}), $\bcG(-\infty)$ and $\bcG(+\infty)$ store the scattering properties of the potential for the left- and right-incident waves, respectively.

Substituting (\ref{GG=}) and (\ref{reciprocity}) in (\ref{M-GG}), we arrive at the well-known formula \cite{sanchez,bookchapter}:
    \be
    \bM=\frac{1}{T}\left[\begin{array}{cc}
    T^2-R^lR^r & R^r\\
    -R^l & 1\end{array}\right],
    \label{M=RT}
    \ee
where $T$ labels the common value of $T^l$ and $T^r$. Equations
(\ref{RT}) and the fact that $\det\bM=1$ follow as simple
consequences of (\ref{M=RT}). This provides an alternative
verification of item 1 in the list of properties of the transfer
matrix that we have given in Sec.~1. Ref.~\cite{bookchapter}
outlines the standard derivation of the composition property
(\ref{composition}) of transfer matrices which is the content of
item 2 of this list. In the following we pursue an alternative route
for establishing this property which is in line with the dynamical
formulation of the (short-range) potential scattering
\cite{ap-2014,pra-2014a}.

Consider the two-component wave function,
    \be
    \Psi(x):=\bcF_0(x)^{-1}\left[\begin{array}{c}
    \psi(x)\\[6pt]
    \psi'(x)\end{array}\right]=
    \frac{1}{2}\left[\begin{array}{c}
    e^{-ikx}[\psi(x)-ik^{-1}\psi'(x)]\\[6pt]
    e^{ikx}[\psi(x)+ik^{-1}\psi'(x)]\end{array}\right],
    \label{Psi-def}
    \ee
where $\psi$ is a general solution of the Schr\"odinger equation (\ref{sch-eq}) for the short-range potential $v$. Differentiating both sides of $\Psi$ and making use of (\ref{sch-eq}), we find that $\Psi$ satisfies the time-dependent Schr\"odinger equation,
    \be
    i\Psi'(x)=\sH(x)\Psi(x),
    \label{time-dep}
    \ee
where $x$ plays the role of time, $\sH(x)$ is the non-stationary matrix Hamiltonian,
    \bea
    &&\sH(x):=\frac{v(x)}{2k}\left[\begin{array}{cc}
    1 & e^{-2ikx}\\
    -e^{2ikx} & -1\end{array}\right]
    =\frac{v(x)}{2k}\, e^{-ikx\bsigma_3} \bcK\, e^{ikx\bsigma_3},
    \label{H-def-F}
    \eea
and
    \bea
    &&\bcK:=i\bsigma_2+\bsigma_3=\left[\begin{array}{cc}
    1 & 1\\
    -1 & -1\end{array}\right].
    \label{K-def}
    \eea
Furthermore, according to (\ref{FM-3}) and (\ref{Psi-def}),
    \be
    \Psi(x)\to\left[\begin{array}{c}
    A_\pm\\
    B_\pm\end{array}\right]~~{\rm for}~~x\to\pm\infty.
    \label{Psi-asym}
    \ee
This relation together with (\ref{M-def}) and (\ref{time-dep}) allow
us to identify $\bM$ with $\sU(\infty,-\infty)$, where $\sU(x,x_0)$
is the evolution operator associated with the Hamiltonian $\sH(x)$
and the initial `time' $x_0$, \cite{ap-2014}.\footnote{By
definition, $\sU(x,x_0)$ satisfies $\Psi(x)=\sU(x,x_0)\Psi(x_0)$ for
all $x\in\R$. In particular, $\sU(x_0,x_0)=\bI$.}

Because we can express $\sU(x,x_0)$ as the time-ordered exponential of $\sH(x)$, i.e.,
    \bea
    \sU(x,x_0)&=&\sT\exp\left\{-i\int_{x_0}^x\sH(s)ds\right\}
    \label{U=T-exp}\\
    &=&\bI+\sum_{\ell=1}^\infty(-i)^\ell
    \int_{x_0}^xdx_\ell \int_{x_0}^{x_\ell}dx_{\ell-1}\cdots
    \int_{x_0}^{x_{2}}dx_{1}\sH(x_\ell)\sH(x_{\ell-1})\cdots\sH(x_1),\nn
    \eea
we have
    \be
    \bM=\sU(\infty,-\infty)=\sT\exp\left\{-i\int_{-\infty}^\infty\sH(x)dx\right\}.
    \label{T-exp}
    \ee
The transfer matrix possesses the composition property (\ref{composition}), because evolution operators satisfy the semi-group multiplication rule, $\sU(x_2,x_0)=\sU(x_2,x_1)\sU(x_1,x_0)$ for all $x_0,x_1,x_2\in\R$, and the fact that $\sH(x)$ vanishes for values of $x$ for which $v(x)=0$, \cite{ap-2014,pra-2014a}. Another notable consequence of (\ref{T-exp}) is that because $\sH(x)$ is traceless, $\sU(x,x_0)$ and consequently $\bM$ have a unit determinant.

If we identify $\sH(x)$ with the interaction Hamiltonian for a two-level quantum system, $\bM$ which is equal to $\sU(\infty,-\infty)$ gives the $S$-matrix of this system \cite{weinberg}. Note however that $\sH(x)$ is manifestly non-Hermitian (and non-diagonalizable) even if $v$ is a real-valued potential. In the latter case, it is $\bsigma_3$-pseudo-Hermitian \cite{p123}, i.e., $\sH(x)^\dagger=\bsigma_3\sH(x)\bsigma_3^{-1}$. If $v$ is a complex potential, $\sH(x)$ is $\bsigma_3$-pseudo-normal, i.e., it commutes with its $\bsigma_3$-pseudo-adjoint, $\bH(x)^\sharp:=\bsigma_3^{-1}\sH(x)^\dagger\bsigma_3$, \cite{p123}.

Because $\psi_1$ and $\psi_2$ are solutions of the Schr\"odinger equation (\ref{sch-eq}), the corresponding two-component wave functions, $\Psi_1$ and $\Psi_2$, solve (\ref{time-dep}). In light of (\ref{G=}), this implies that
    \be
    i\bcG'(x)=\sH(x)\bcG(x).
    \label{G-eqn}
    \ee
Equivalently, we have
    \be
    \bcG(x)=\sU(x,x_0)\bcG(x_0),
    \label{G=UG}
    \ee
which gives rise to
    \be
    \sU(x,x_0)=\bcG(x)\bcG(x_0)^{-1}.
    \label{identity}
    \ee
Letting $x_0\to-\infty$ and $x\to+\infty$ in this relation and using (\ref{M-GG}), we recover (\ref{T-exp}). Note also that Eqs.~(\ref{G-eqn}) -- (\ref{identity}) are valid for arbitrary choices of the linearly-independent solutions $\psi_1$ and $\psi_2$ of the Schr\"odinger equation (\ref{sch-eq}); they need not coincide with the scattering solutions $\psi_l$ and $\psi_r$.

\section{Generalization to long-range potentials}
\label{S3}

Let $\alpha_0$ be a real number and $\cC_{\alpha>\alpha_0}$ denote the class of potentials satisfying (\ref{condi-0}) for some $\alpha>\alpha_0$, so that $\cC_{\alpha>1}$ specifies the set of short-range potentials.\footnote{We can extend the definition of $\cC_\alpha$ and notions of short- and long-range potentials to $d$ dimensions by identifying the independent variable $x$ in (\ref{condi-0}) with an element of $\R^d$.} The scattering theory of the latter is a well-established mathematical discipline \cite{reed-simon}. Extending this theory to encompass long-range potentials has been an active area of research since the 1960's \cite{dollard,lavine,enss,derezinskia,christ-kiselev-1998,christ-kiselev-2002}. This has primarily been motivated by the indisputable physical importance of long-range interactions, such as the Coulomb interaction. The recent discovery of the application of complex long-range potentials in realizing unidirectional invisibility for all frequencies \cite{longhi-2015,longhi-2016} has also drawn attention to the scattering theory of complex long-range potentials.

For real-valued potentials $v$ belonging to $\cC_{\alpha>1/2}$, the absolutely continuous spectrum of the Schr\"odinger operator $-\partial_x^2+v(x)$ coincides with $[0,\infty)$ and its generalized eigenfunctions have the asymptotic WKB form \cite{christ-kiselev-1998,christ-kiselev-2002}:
        \be
        \psi(x)\to \brv A_\pm e^{iS(x)}+\brv B_\pm e^{-iS(x)}
        ~~~{\rm as}~~~x\to\pm\infty,
        \label{e2}
        \ee
where $\brv A_\pm$ and $\brv B_\pm$ are constant coefficients,
        \bea
        S(x)&:=&kx+\varsigma(x)=
        kx\left[1+\frac{V(x)}{2k^2}\right],
        \label{e3}\\
        \varsigma(x)&:=&-\frac{1}{2k}\int_0^x v(s) ds,
        \label{f-def}
        \eea
and $V(x):=-\frac{1}{x}\int_0^x v(s) ds$. It is not difficult to
show that $V$ belongs to $\cC_{\alpha>1/2}$. In particular, $S$
tends to an increasing function of $x$ as $x\to\pm\infty$. This in
turn allows for identifying $e^{iS(x)}$ and $e^{-iS(x)}$
respectively with asymptotic right- and left-going waves, and
suggests defining the transfer matrix of $v$ as the $2\times 2$
matrix $\brv\bM$ that satisfies,
    \be
        \left[\begin{array}{cc}
        \brv A_+\\ \brv B_+\end{array}\right]
        =\brv \bM\left[\begin{array}{cc}
        \brv A_-\\\brv B_-\end{array}\right],
        \label{M-def-brv}
     \ee
and is independent of $\brv A_-$ and $\brv B_-$.

If we identify the scattering solutions, $\brv\psi_l$ and $\brv\psi_r$, of the Schr\"odinger equation with those fulfilling the asymptotic boundary conditions:
    \bea
    \brv\psi_l(x)&\to&\left\{\begin{array}{ccc}
    e^{iS(x)} + \brv R^l e^{-iS(x)} &{\rm for}&x\to-\infty,\\
    \brv T^l e^{iS(x)} &{\rm for}&x\to+\infty,\end{array}\right.
    \label{psi-left}\\[6pt]
    \brv\psi_r(x)&\to&\left\{\begin{array}{ccc}
    \brv T^r e^{-iS(x)} & {\rm for}&x\to-\infty,\\
    e^{-iS(x)} + {\brv R}^r e^{iS(x)} & {\rm for}&x \to +\infty,
    \end{array}\right.
    \label{psi-right}
    \eea
and identify the reflection and transmission amplitudes of the potential with the coefficients $\brv R^{l/r}$ and $\brv T^{l/r}$ appearing in these relations, we are led to the following analog of (\ref{RT}).
     \begin{align}
    &\brv R^l=-\frac{\brv M_{21}}{\brv M_{22}},
    &&\brv R^r=\frac{\brv M_{12}}{\brv M_{22}},
    &&\brv T^{l/r}=\brv T:=\frac{1}{\brv M_{22}}.
    \label{RT-brv}
    \end{align}

A direct implementation of this prescription to short-range real
potentials shows that the transfer matrix $\brv\bM$ and the
reflection and transmission amplitudes, $\brv R^{l/r}$ and $\brv
T^{l/r}$, differ from the standard transfer matrix $\bM$ and the
reflection and transmission amplitudes, $R^{l/r}$ and $T^{l/r}$.
This is simply because for a short-range potential,
    \be
    \vartheta^\pm:=\varsigma(\pm\infty):=\lim_{x\to\pm\infty}\varsigma(x)
    \label{vartheta}
    \ee
are finite but not necessarily zero. As a result, (\ref{e2}) would agree with (\ref{e2-SR}) provided that
    \begin{align}
    &\brv A_\pm= e^{-i\vartheta^\pm} A_\pm,
    &&\brv B_\pm=e^{i\vartheta^\pm} B_\pm.
    \label{AB=AB}
    \end{align}
In view of (\ref{M-def}), (\ref{M-def-brv}), and (\ref{AB=AB}), $\bM$ and $\brv\bM$ are related via
    \be
    \bM=e^{i\vartheta^+\bsigma_3}\brv\bM\,e^{-i\vartheta^-\bsigma_3}.
    \label{M=M-brv}
    \ee
This equation together with (\ref{RT}) and (\ref{RT-brv}) imply
    \begin{align}
    &R^l=e^{-2i\vartheta^-}\brv R^l,
    &&R^r=e^{2i\vartheta^+}\brv R^r,
    &&T=e^{i(\vartheta^+-\vartheta^-)}\brv T.
    \label{RRT=trans1}
    \end{align}

We can similarly introduce a transfer matrix for complex-valued potentials belonging to $\cC_{\alpha>1/2}$ provided that $|e^{\pm i S(x)}|^2$ tend to finite values as $x\to\pm\infty$. This restricts the imaginary part of the potential to be short-range. In what follows we confine our discussion to this class of complex long-range potentials, i.e., consider complex-valued potentials $v$ such that
    \begin{align}
    &\RE(v)\in\cC_{\alpha>1/2},
    &&\IM(v)\in\cC_{\alpha>1},
    \label{condi-complex}
    \end{align}
where `$\RE$' and `$\IM$' stand for the real and imaginary parts of their argument, respectively.

Let,
    \[\vartheta_{i}^{\pm}:=\lim_{x\to\pm\infty}\IM[\varsigma(x)].\]
Then in view of (\ref{condi-complex}), $\vartheta_{i}^{\pm}$ are real numbers, and $|e^{i S(x)}|\to e^{-\vartheta_{i}^{\pm}}$ for $x\to\pm\infty$. Therefore, the $e^{\pm i S(x)}$, that appear in (\ref{psi-left}) and (\ref{psi-right}) are not generally unimodular. This would be in conflict with the identification of $|\brv R^{l/r}|^2$ and $|\brv T^{l/r}|^2$ with the reflection and transmission coefficients, because these coefficients are respectively defined as the ratio of the intensity of the reflection and transmitted waves to the intensity of the incident wave \cite{griffiths-book}. To avoid this conflict, we introduce
    \be
     S_\pm(x):=S(x)-i\vartheta_{i}^{\pm},
    \label{S-pm}
    \ee
and express the asymptotic expression for the scattering solutions of the Schr\"odinger equation in the form
    \bea
    \psi_l(x)&\to&\left\{\begin{array}{ccc}
    e^{iS_-(x)} + \cR^l e^{-iS_-(x)} &{\rm for}&x\to-\infty,\\
    \cT^l e^{iS_+(x)} &{\rm for}&x\to+\infty,\end{array}\right.
    \label{psi-left-brv3}\\[6pt]
    \psi_r(x)&\to&\left\{\begin{array}{ccc}
    \cT^r e^{-iS_-(x)} & {\rm for}&x\to-\infty,\\
    e^{-iS_+(x)} + {\cR}^r e^{iS_+(x)} & {\rm for}&x \to +\infty,
    \end{array}\right.
    \label{psi-right-brv4}
    \eea
where $\cR^{l/r}$ and $\cT^{l/r}$ are respectively the left/right reflection and transmission amplitudes.

Relations (\ref{psi-left-brv3}) and (\ref{psi-right-brv4}) suggest that we express the asymptotic form of the general solution of the Schr\"odinger equation (\ref{sch-eq}) for complex potentials subject to the conditions (\ref{condi-complex}) as
        \be
        \psi(x)\to \cA_\pm e^{i S_\pm(x)}+\cB_\pm e^{-i S_\pm(x)}
        ~~~{\rm as}~~~x\to\pm\infty,
        \label{e2-LR-brv-c}
        \ee
where $\cA_\pm$ and $\cB_\pm$ are constant coefficients. Comparing (\ref{e2}) and (\ref{e2-LR-brv-c}), we observe that
    \begin{align}
    \cA_\pm= e^{-\vartheta_{i}^{\pm}} \brv A_\pm,
    && \cB_\pm=e^{\vartheta_{i}^{\pm}} \brv B_\pm.
    \label{rescale}
    \end{align}

We identify the transfer matrix for this class of complex potentials with the $2\times 2$ matrix $\bcM$ satisfying
    \be
        \left[\begin{array}{cc}
        \cA_+\\ \cB_+\end{array}\right]
        =\bcM\left[\begin{array}{cc}
        \cA_-\\ \cB_-\end{array}\right].
        \label{M-def-brv-c}
     \ee
Again, we can relate the reflection and tranmission amplitudes, $\cR^{l/r}$ and $\cT^{l/r}$, to the entries of $\bcM$;
    \begin{align}
    &\cR^l=-\frac{\cM_{21}}{\cM_{22}},
    &&\cR^r=\frac{\cM_{12}}{\cM_{22}},
    &&\cT^{l/r}=\cT:=\frac{1}{\cM_{22}}.
    \label{RT-brv-c}
    \end{align}
With the help of (\ref{M-def-brv}), (\ref{rescale}), and (\ref{M-def-brv-c}),  we can express $\bcM$ in terms of $\brv\bM$ according to
    \be
    \bcM=e^{-\vartheta_{i}^+\bsigma_3}\brv\bM\, e^{\vartheta_{i}^-\bsigma_3}.
    \label{brv-M=3}
    \ee
This equation together with (\ref{RT-brv}) and (\ref{RT-brv-c}) imply
    \begin{align}
        &\cR^l=-\frac{e^{2\vartheta_i^-}\brv M_{21}}{\brv M_{22}}=
    e^{2\vartheta_i^-}\brv R^l,
        &&\cR^r=\frac{e^{-2\vartheta_i^+}\brv M_{12}}{\brv M_{22}}=
    e^{-2\vartheta_i^+}\brv R^r,
        &&\cT:=\frac{e^{\vartheta_i^--\vartheta_i^+}}{\brv M_{22}}=
    e^{\vartheta_i^--\vartheta_i^+}\brv T.
        \label{RT-brv-c2}
        \end{align}
It is also not difficult to show that the standard  transfer matrix
$\bM$ for short-range complex potentials is given by
    \be
    \bM=e^{i\vartheta^+\bsigma_3}\brv\bM\, e^{-i\vartheta^-\bsigma_3}=
    e^{i\vartheta_{r}^+\bsigma_3}\bcM\, e^{-i\vartheta_{r}^-\bsigma_3},
    \label{brv-M=SR}
    \ee
where $\vartheta^\pm_r:=\RE(\vartheta^\pm)$.

Next, we explore the relationship between the transfer matrix
$\brv\bM$ and the classical notion of a fundamental matrix of a
second order ordinary differential equation. To do this, we
introduce:
    \bea
    \brv\bcF_0(x)&:=&\left[\begin{array}{cc}
    e^{iS(x)}&e^{-iS(x)}\\
    ik e^{iS(x)}&-ike^{-iS(x)}\end{array}\right]=
    \bcF_0(0)\,e^{iS(x)\sigma_3},
    \label{FM-0-LR}\\
    \brv\Psi(x)&:=&\brv\bcF_0(x)^{-1}
    \left[\begin{array}{c}
    \psi(x)\\
    \psi'(x)\end{array}\right]=
    \frac{1}{2}\left[\begin{array}{c}
    e^{-iS(x)}\{\psi(x)-ik^{-1}\psi'(x)\}\\
    e^{iS(x)}\{\psi(x)+ik^{-1}\psi'(x)\}\end{array}\right],
    \label{Psi-def-brv}\\
    \brv\Psi_j(x)&:=&\brv\bcF_0(x)^{-1}
    \left[\begin{array}{c}
    \psi_j(x)\\[6pt]
    \psi_j'(x)\end{array}\right],
    \label{2-comp-LR}\\
    \brv\bcG(x)&:=&\brv\bcF_0(x)^{-1}\bcF(x)=[\:\brv\Psi_1(x)~~\brv\Psi_2(x)\:],
    \label{bcG-brv}
    \eea
where $\psi$ is the general solution of the Schr\"odinger equation
(\ref{sch-eq}), $\psi_j$ with $j\in\{1,2\}$ are linearly-independent
solutions of this equation, and $\bcF(x)$ is
the corresponding fundamental matrix (\ref{FM-def}). We can use
(\ref{e2}) to show that
    \be
    \brv\Psi(x)\to\left[\begin{array}{c}
    \brv A_\pm\\
    \brv B_\pm\end{array}\right]~~{\rm for}~~x\to\pm\infty.
    \label{Psi-asym-brv}
    \ee
This relation together with (\ref{FM-2}), (\ref{M-def-brv}),
(\ref{Psi-def-brv}), (\ref{2-comp-LR}), and (\ref{bcG-brv}) imply
    \bea
    \brv\bM&=&\brv\bcG(+\infty)\brv\bcG(-\infty)^{-1}.
    \label{M=GG-brv}
    \eea
Substituting this in (\ref{brv-M=3}), we find
    \be
    \bcM=e^{-\vartheta_i^+\bsigma_3}\brv\bcG(+\infty)\brv\bcG(-\infty)^{-1}
    \, e^{\vartheta_i^-\bsigma_3}.
    \label{brv-M=}
    \ee

If we respectively identify $\psi_1$ and $\psi_2$  with the
scattering solutions (\ref{psi-left}) and (\ref{psi-right}), we
obtain (\ref{GG=}) -- (\ref{M=RT}) with $\cG(\pm\infty)$, $R^{l/r}$,
$T^{l/r}$, and $\bM$ replaced with $\brv\cG(\pm\infty)$, $\brv
R^{l/r}$, $\brv T^{l/r}$, and $\brv \bM$. Together with
(\ref{brv-M=3}), this provides an alternative derivation of
(\ref{RT-brv-c}) and shows that the transfer matrices $\brv\bM$ and
$\bcM$ share Property 1 of the transfer matrix of the short-range
potentials that we have listed in Sec.~1. The same holds for
Property 2. As we show in the sequel, this follows from the fact
that $\brv\bM$ coincides with the $S$-matrix of an associated
effective two-level quantum system.

In order to derive the composition property of $\brv\bM$, we first
use (\ref{sch-eq}), (\ref{e2}), (\ref{e3}), and
(\ref{Psi-def-brv}) to show that $\brv\Psi$ satisfies
    \begin{align}
    &i\brv\Psi'(x)=\brv\sH(x)\brv\Psi(x),
    \label{time-dep-brv}
    \end{align}
where
    \be
    \brv\sH(x):=\frac{v(x)}{2k}\left[\begin{array}{cc}
    0 & e^{-2iS(x)}\\
    -e^{2iS(x)} & 0\end{array}\right]
    =\frac{iv(x)}{2k}\, e^{-iS(x)\bsigma_3}
    \bsigma_2 e^{iS(x)\bsigma_3}.
    \label{H-def}
    \ee
In view of (\ref{M-def-brv}), (\ref{Psi-asym-brv}), and
(\ref{time-dep-brv}),
    \be
    \brv\bM=\brv\sU(\infty,-\infty)=\sT\exp\left\{-i\int_{-\infty}^\infty\brv\sH(x)dx\right\},
    \label{T-exp-brv}
    \ee
where $\brv\sU(x,x_0)$ is the evolution operator for the Hamiltonian $\brv\sH(x)$ and the initial `time' $x_0$, i.e.,
    \be
    \brv\sU(x,x_0)=\sT\exp\left\{-i\int_{x_0}^x\brv\sH(s)ds\right\}.
    \label{sU-brv}
    \ee
We can also establish (\ref{T-exp-brv}) using (\ref{M=GG-brv}) and
    \be
    \brv\sU(x,x_0)=\brv\bcG(x)\brv\bcG(x_0)^{-1},
    \label{U=GG-brv}
    \ee
which follows from (\ref{bcG-brv}) and (\ref{time-dep-brv}).

Now, consider the truncated potentials $v_j$ given by (\ref{vj-def}), and let $\brv\bM_j$ and $\bcM_j$ be the analogs of the transfer matrices $\brv\bM$ and $\bcM$ for these potentials. Then, we can use (\ref{brv-M=}), (\ref{T-exp-brv}), the semi-group multiplication rule for the evolution operators, and the vanishing of $\brv\sH(x)$ for all $x\in\R$ at which $v(x)=0$ to establish
    \be
    \brv\bM=\brv\bM_n\brv\bM_{n-1}\cdots\brv\bM_1.
    \label{composition-brv}
    \ee
Furthermore, because
    \begin{align*}
    &\lim_{x\to+\infty}v_1(x)=\lim_{x\to-\infty}v_n(x)=0,
    &&\lim_{x\to\pm\infty}v_j(x)=0~~{\rm for}~~j\in\{2,3,\cdots,n-1\},
    \end{align*}
Eq.~(\ref{brv-M=3}) implies that
    \begin{align*}
    &\bcM_1=\brv\bM_1e^{\vartheta_{i}^-\bsigma_3},
    &&\bcM_n=e^{-\vartheta_{i}^+\bsigma_3}\brv\bM_n,
    &&\bcM_j=\brv\bM_j~~{\rm for}~~j\in\{2,3,\cdots,n-1\}.
    \end{align*}
Substituting these relations and (\ref{brv-M=3}) in (\ref{composition-brv}), we arrive at the composition property of the transfer matrix $\bcM$, namely
    \be
    \bcM=\bcM_n\bcM_{n-1}\cdots\bcM_1.
    \label{composition-brv-c}
    \ee
Therefore, $\brv\bM$ and $\bcM$  share the composition property of
the  well-known transfer materix $\bM$ for the short-range
potentials.

According to (\ref{T-exp-brv}), $\brv\sH(x)$ is the Hamiltonian
operator  for an effective two-level quantum system whose $S$-matrix
yields the transfer matrix $\brv\bM$ of $v$. Similarly to $\sH(x)$,
this operator is $\bsigma_3$-pseudo-Hermitian whenever $v$ is
real-valued, and $\bsigma_3$-pseudo-normal
otherwise.\footnote{The main difference between $\sH(x)$ and
$\brv\sH(x)$ is that the latter is diagonalizable.} It is also
traceless which implies $\det\brv\bM=1$. This equation together with
(\ref{brv-M=}) and $\det e^{\pm\vartheta_i^\pm\bsigma_3}=1$ lead to
another proof of the fact that $\det\bcM=1$.

The argument leading to (\ref{T-exp-brv}) is clearly applicable to
short-range potentials. For a short-range potential,
$S(x)=kx+\varsigma(x)\to kx+\vartheta^\pm$ as $x\to\pm\infty$.
According to (\ref{FM-0}), (\ref{Psi-def}), (\ref{FM-0-LR}), and
(\ref{Psi-def-brv}), this implies
    \bea
    &&\brv\Psi(x)=e^{-i \varsigma(x)\bsigma_3}\Psi(x),
    \label{trans1}\\
    &&\brv\sU(x,x_0)=e^{-i \varsigma(x)\bsigma_3}\sU(x,x_0)e^{i
    \varsigma(x_0)\bsigma_3},
    \label{trans3}
    \eea
where we have also employed (\ref{identity}) and (\ref{U=GG-brv}).
Taking $x\to+\infty$ and $x_0\to-\infty$ in (\ref{trans3}) and
making use of (\ref{T-exp}) and (\ref{T-exp-brv}) we recover
(\ref{brv-M=SR}).

\section{Long-range potentials as short-range perturbations of\\ exactly solvable potentials}
\label{S4}

Consider a long-range potential $v$ fulfilling (\ref{condi-complex}). For every positive real number $a$ of our choice, we can dissect the real line into the intervals:
    \begin{align*}
    & I_{-}:=(-\infty,-a], && I_0:=(-a,a), && I_{+}:=[a,+\infty),
    \end{align*}
introduce the potentials
    \be
    v_j(x):=\left\{\begin{array}{ccc}
        v(x) & {\rm for} & x\in I_j,\\
        0 & {\rm for} & x\notin I_j,\end{array}\right.
        \label{vj-def}
        \ee
with $j\in\{-,0,+\}$, so that
    \be
    v=v_-+v_0+v_+,
    \label{decompose-v}
    \ee
and express the transfer matrix $\brv\bM$ of $v$ in the form
$\brv\bM=\brv\bM_{+}\brv\bM_0\brv\bM_{-}$, where $\brv\bM_j$ is the
transfer matrix of $v_j$. Clearly, $v_0$ is a short-range potential.
Therefore, in dealing with the difficulties associated with the long
range of $v$, we can focus our attention to $v_{\pm}$. Because under
a reflection (parity) transformation $v_{-}$ is mapped to a
potential with the same structure as $v_{+}$, we confine our
investigation to long-range potentials of the form $v_{+}$, i.e.,
those supported in $I_{+}$. In the following, we derive a
decomposition of $v_{+}$ into the sum of a short-range potential $u$
and an exactly solvable long-range potential $w$. This is of
interest, because for sufficiently large values of $a$, we can treat
$v_+$ as a perturbation of $w$.

Let $\epsilon$ be a real number such that $0<\epsilon<1$. Because $v$ belongs to $\cC_{\alpha>1/2}$, for every $k$ there is a positive real number $a_0$ such that
    \be
    \left|1-\frac{v(x)}{2k^2}\right|\geq\epsilon~~{\rm for~all}~~x\geq a_0.
    \label{condi-1}
    \ee
In the following, we choose $a\geq a_0$ and introduce the functions $f_{\pm}:[a,\infty)\to\C$ and $\psi_{\pm}:\R\to\C$ according to
    \bea
    f_{\pm}(x)&:=&\frac{e^{\pm i S(x)}}{\sqrt{1-v(x)/2k^2}}~~{\rm for}~~x\geq a,
    \label{fk=}\\[6pt]
    \psi_{\pm}(x)&:=&\left\{\begin{array}{ccc}
    f_{\pm}(x)& {\rm for} & x\geq a,\\
    \fa_\pm e^{ikx} + \fb_\pm e^{-ikx}& {\rm for} & x< a,\end{array}\right.
    \label{psi-pm=}
    \eea
where $S$ is given by (\ref{e3}), and $\fa_\pm$ and $\fb_\pm$ are complex coefficients that render $\psi_{\pm}$ differentiable at $x=a$, i.e.,
    \bea
    \fa_\pm&=&\frac{e^{-ika}}{2}\left[f_\pm(a)-if'_\pm(a)/k\right],
    \label{fa-def}\\
    \fb_\pm&=&\frac{e^{ika}}{2}\left[f_\pm(a)+if'_\pm(a)/k\right].
    \label{fb-def}
    \eea

It is not difficult to check that $\psi_{\pm}$ are solutions of the time-independent Schr\"odinger equation (\ref{sch-eq}) for a potential of the form,
    \be
    w(x):=\left\{\begin{array}{ccc}
    v(x)-u(x) &{\rm for} & x\geq a,\\
    0 &{\rm for} & x< a,\end{array}\right.
    \label{wk-def}
    \ee
where
    \bea
    u(x)&:=&\left\{\begin{array}{cc}
    \displaystyle
    \frac{1}{4k^2}\left[v(x)^2-\frac{3v'(x)^2}{4k^2\tau(x)^2}-\frac{v''(x)}{\tau(x)}\right]
    &{\rm for}~~x\geq a,\\[9pt]
    0 &{\rm for}~~x< a,\end{array}\right.
    \label{uk-def}\\
    \tau(x)&:=&1-\frac{v(x)}{2k^2}.
    \label{tau-def}
    \eea
Because $\psi_{\pm}$ are linearly independent, every solution $\psi$ of the Schr\"odinger equation (\ref{sch-eq}) for the potential $w$ is a linear combination of $\psi_{\pm}$; there are complex coefficients $A$ and $B$ such that
    \be
    \psi(x)=A\, \psi_{+}(x)+B\,\psi_{-}(x).
    \label{gen-sol}
    \ee
We can use this relation to determine the transfer matrix of $w$. To
this end, we first introduce\footnote{Because $u$ is a short-range
potential, $\varsigma_u(\infty):=\lim_{x\to\infty}\varsigma_u(x)$
exists.}:
    \begin{align}
    &\varsigma_u(x):=-\frac{1}{2k}\int_0^x u(s)ds,
    &&S_w(x):=S(x)-\varsigma_u(x)=kx-\frac{1}{2k}\int_0^x w(s)ds,
    \label{Sw-def}\\
    & \brv A_+:= A \,e^{i\varsigma_u(\infty)},
    && \brv B_+:=  B \,e^{-i\varsigma_u(\infty)},
    \label{AB=AB-brv-2}
    \end{align}
and use (\ref{fk=}), (\ref{psi-pm=}), (\ref{gen-sol}), and
(\ref{Sw-def}) to show that
    \be
    \psi(x)\to \brv A_+\, e^{iS_w(x)}+\brv B_+\,e^{-iS_w(x)}~~{\rm for}~~x\to+\infty.
    \label{gen-sol-asym-p}
    \ee
Moreover, for $x<a$,
    \bea
    \psi(x)&=&(\fa_+ A_++\fa_- B_+)e^{ikx}+
    (\fb_+ A_++\fb_- B_+)e^{-ikx}\nn\\
    &=&(\fa_+e^{-i\varsigma_u(\infty)}\brv A_++\fa_-e^{i\varsigma_u(\infty)}\brv B_+)e^{iS_w(x)}+\nn\\
    &&(\fb_+e^{-i\varsigma_u(\infty)}\brv A_++\fb_-e^{i\varsigma_u(\infty)}\brv B_+)e^{-iS_w(x)},
    \label{gen-sol-asym-m}
    \eea
where we have made use of (\ref{AB=AB-brv-2}) and the fact that $S_w(x)=kx$ for $x<a$.

Next, we observe that because (\ref{gen-sol-asym-m}) holds for $x\to-\infty$, we can identify the
coefficients of $e^{iS_w(x)}$ and $e^{-iS_w(x)}$ on the right-hand side of (\ref{gen-sol-asym-m}) with
$\brv A_-$ and $\brv B_-$. This yields a pair of linear equations
for $\brv A_+$ and $\brv B_+$. Expressing the solution of these
equations in the form (\ref{M-def-brv}), we find the following
formula for the transfer matrix $\brv \bM$ of the potential $w$,
which we label by $\brv\bM_w$.
    \bea
    \brv \bM_w&=&\left[\begin{array}{cc}
    \fa_+ e^{-i\varsigma_u(\infty)}& \fa_-e^{i\varsigma_u(\infty)}\\
    \fb_+ e^{-i\varsigma_u(\infty)}& \fb_-e^{i\varsigma_u(\infty)}\end{array}\right]^{-1}=
    \left[\brv\bcG_w(a)e^{-i\varsigma_u(\infty)\bsigma_3}\right]^{-1}=
    e^{i\varsigma_u(\infty)\bsigma_3}\brv\bcG_w(a)^{-1}.
    \label{Mw=}
    \eea
where $\brv\bcG_w$ is the matrix-valued function (\ref{bcG-brv})
associated with the potential $w$, and we have employed the
identity,
    \be
    \brv\bcG_w(a)=\left[\begin{array}{cc}
        \fa_+ & \fa_-\\
        \fb_+ & \fb_-\end{array}\right],
        \label{brv-G-a=}
    \ee
which we obtain by setting $\psi_1=\psi_+$ and $\psi_2=\psi_-$ in (\ref{2-comp-LR}) and using the resulting equation together with (\ref{bcG-brv})  and (\ref{psi-pm=}) to compute $\brv\bcG_w(a)$. Because $\brv \bM_w$ has a unit determinant, (\ref{Mw=}) implies
    \be
    \fa_+\fb_--\fa_-\fb_+=1.
    \label{ab-ba}
    \ee
We can indeed verify this relation by exploiting the fact that the
computation of the Wronskian of the solutions $\psi_\pm$ at $x=0$
and in the limit $x\to\infty$ gives the same result. Employing
(\ref{psi-pm=}) to perform this calculation, we respectively find
$2ik(\fa_+\fb_--\fa_-\fb_+)$ and $2ik$. Hence (\ref{ab-ba}) holds.
Using this equation in Eqs.~(\ref{Mw=}), we have
     \bea
     \brv \bM_w&=&\left[\begin{array}{cc}
     \fb_- e^{i\varsigma_u(\infty)}& -\fa_-e^{i\varsigma_u(\infty)}\\
    -\fb_+ e^{-i\varsigma_u(\infty)}& \fa_+e^{-i\varsigma_u(\infty)}\end{array}\right].
    \label{Mw=2}
    \eea

Next, we recall that, in light of (\ref{vj-def}) and (\ref{wk-def}),
    \be
    v_{+}(x)=\left\{\begin{array}{ccc}
    u(x)+w(x)&{\rm for}& x\geq a,\\
    0&{\rm for}& x<a.\end{array}\right.
    \label{v=uw}
    \ee
If there is some $\alpha>1/2$ such that $v(x)\propto x^{-\alpha}$ as $x\to +\infty$, then (\ref{uk-def}) implies that in this limit $u(x)\propto x^{-\alpha'}$ for some $\alpha'>1$, i.e.,  $u$ is a short-range potential. According to (\ref{uk-def}), this is generally true, for potentials $v$ of class $\cC_{\alpha>1/2}$ such that $v'$ also belongs to $\cC_{\alpha>1/2}$ and $v''$ is a short-range potential. Under these conditions $v_{+}$ is the sum of a short-range potential $u$ and an exactly solvable long-range potential $w$.

The constructions leading to (\ref{v=uw}) are clearly valid for every $a\geq a_0$. This together with the fact that for larger values of $a$ we can treat $u$ as a small perturbation of $w$ suggest using first-order perturbation theory to compute the transfer matrix $\brv\bM$ of $v_+$, which we denote by $\brv\bM_+$.

Let $\Psi_q$, $\sH_q(x)$, and $\sU_q(x,x_0)$ respectively stand for the two-component wave function (\ref{Psi-def}), the Hamiltonian (\ref{H-def-F}), and the evolution operator (\ref{U=T-exp}) for the potential $q\in\{v_{+},w,u\}$. Then, the two-component wave function $\Phi$ defined by,
    \be
    \Phi(x):=\sU_w(x,a)^{-1}\Psi_{v_{+}}(x),
    \label{Phi-def}
    \ee
satisfies $i\Phi'(x)=\bH(x)\Phi(x)$ for
    \bea
    \bH(x)&:=&\sU_w(x,a)^{-1}\sH_u(x)\sU_w(x,a)
    \label{bH-def}\\
    &=&\frac{u(x)}{2k}\:\sU_w(x,a)^{-1} e^{-ikx\bsigma_3}\bcK\:
    e^{ikx\bsigma_3}\sU_w(x,a).\nn
    \eea
In other words,
    \be
    \Phi(x)=\bU(x,x_0)\Phi(x_0),
    \label{Phi-U-Phi}
    \ee
where
    \be
    \bU(x,x_0):=\sT\exp\left\{-i\int_{x_0}^x \bH(s)ds\right\}.
    \label{nU-def}
    \ee

If we respectively denote the two-component wave function (\ref{Psi-def-brv}), the Hamiltonian (\ref{H-def}), and the evolution operator (\ref{sU-brv}) for the potential $q$ by $\brv\Psi_q$, $\brv\sH_q(x)$, and $\brv\sU_q(x,x_0)$, with the help of (\ref{trans1}) and (\ref{trans3}), we can express (\ref{Phi-def}) in the form
    \be
    \Phi(x)=\brv\sU_w(x,a)^{-1}e^{i\varsigma_u(x)\bsigma_3}
    \brv\sU_{v_{+}}(x,x_0)\brv\Psi_{v_{+}}(x_0),
    \label{Phi-Psi}
    \ee
where $\varsigma_q(x):=-\frac{1}{2k}\int_0^x q(s)ds$, and we have
benefitted from the identities: $\varsigma_w(a)=0$ and
$\varsigma_{v_{+}}-\varsigma_w=\varsigma_u$. For $x=x_0$,
(\ref{Phi-Psi}) gives $\Phi(x_0)=\brv\sU_w(x_0,a)^{-1}
e^{i\varsigma_{u}(x_0)\bsigma_3}\brv\Psi_{v_{+}}(x_0)$. Solving this
equation for $\brv\Psi_{v_{+}}(x_0)$ and inserting the result in
(\ref{Phi-Psi}), we recover (\ref{Phi-U-Phi}) with
    \[\bU(x,x_0)=\brv\sU_w(x,a)^{-1}e^{i\varsigma_u(x)\bsigma_3}
    \brv\sU_{v_{+}}(x,x_0)e^{-i\varsigma_{u}(x_0)\bsigma_3}\brv\sU_w(x_0,a).\]
    According to this equation,
    \be
    \brv\sU_{v_{+}}(x,x_0)=e^{-i\varsigma_u(x)\bsigma_3}\brv\sU_w(x,a)
    \bU(x,x_0)\brv\sU_w(x_0,a)^{-1}
    e^{i\varsigma_{u}(x_0)\bsigma_3}.
    \label{U=xUx}
    \ee

Next, we recall that $\varsigma_u(a)=0$ and the transfer matrices of
$v_{+}$ and $w$ are respectively given by
    \begin{align}
    &\brv \bM_{{+}}=\brv\sU_{v_{+}}(\infty,-\infty)=\brv\sU_{v_{+}}(\infty,a),
    &&\brv \bM_{w}=\brv\sU_{w}(\infty,-\infty)=\brv\sU_{w}(\infty,a).
    \end{align}
In view of these observations, letting $x_0=a$ and $x\to\infty$ in (\ref{U=xUx}) and making use of (\ref{Mw=}), we arrive at
    \be
    \brv \bM_{{+}}=e^{-i\varsigma_u(\infty)\bsigma_3}\brv \bM_{w}\bU(\infty,a)=
    \brv\bcG_{w}(a)^{-1}\bU(\infty,a).
    \label{M=MU}
    \ee
Given that we have an explicit formula for $\brv\bcG_{w}(a)$, namely (\ref{brv-G-a=}), this equation reduces the solution of the scattering problem for $v_{+}$ to that of the determination of $\bU(\infty,a)$.

\section{Perturbative evaluation of the transfer matrix}
\label{S5}

Because $u$ is a short-range potential, for sufficiently large values of $a$, we can find positive numbers $\gamma$ and $\delta$ such that $|u(x)|\leq \gamma  k^{1-\delta} x^{-(1+\delta)}$ for $x\geq a$. Hence,
    \be
    \int_a^\infty |u(x)|dx\leq
    \frac{\gamma k}{\delta (ak)^\delta}.
    \label{bound-2}
    \ee
This relation together with the expression  (\ref{bH-def}) for the Hamiltonian $\bH(x)$ and the Dyson series expansion of $\bU(\infty,a)$, i.e.,
    \be
    \bU(\infty,a)= \bI+\sum_{\ell=1}^\infty(-i)^\ell
    \int_{a}^\infty dx_\ell \int_{a}^{x_\ell}dx_{\ell-1}\cdots
    \int_{a}^{x_{2}}dx_{1}\bH(x_\ell)\bH(x_{\ell-1})\cdots\bH(x_1),
    \label{dyson}
    \ee
suggest the possibility of devising a perturbative method of computing $\bU(\infty,a)$ that involves the truncation of its Dyson series. Retaining the first $n+1$ terms of this series, we obtain an $n$-th order perturbative expression for $\bU(\infty,a)$ with $(ak)^{-\delta}$ playing the role of the perturbation parameter.

Consider the fundamental matrix, $\bcF_w(x):=\left[\begin{array}{cc}
    \psi_+(x) & \psi_-(x)\\
    \psi_+'(x) & \psi_-'(x)\end{array}\right]$,  where $\psi_\pm$ are the solutions (\ref{psi-pm=}) of the Schr\"odinger equation (\ref{sch-eq}) for the potential $w$. Then according to (\ref{bcG-brv}), (\ref{U=GG-brv}), (\ref{trans3}), and (\ref{psi-pm=}), for all $x>a$,
    \bea
    \sU_w(x,a)&=&e^{i\varsigma_w(x)\bsigma_3}\brv\bcG_w(x)\brv\bcG_w(a)^{-1},
    \label{Uw-GG}\\
    \brv\bcG_w(x)&=&
    e^{-iS(x)\bsigma_3}\bG(x) e^{iS(x)\bsigma_3},
    \label{G-bound}
    \eea
where
    \begin{align}
    &\bG(x):=\mu_+(x)\bI+\mu_-(x)\bsigma_1+i\nu(x)\bcK,\\
    &\mu_\pm(x):=\frac{1\pm \tau(x)}{2\sqrt{\tau(x)}},
    ~~~~~\nu(x):=-\frac{v'(x)}{8k^3\sqrt{\tau(x)^3}},
    \end{align}
$\tau$ is the function defined by (\ref{tau-def}), and we have made
use of the fact that $\varsigma_w(a)=0$. Because the imaginary part
of $v_{+}$ is a short-range potential, $e^{\pm iS(x)}$, $e^{\pm
i\varsigma_w(x)}$, and consequently the entries of $\brv\bcG_w(x)$
and $\sU_w(x,a)$ are bounded functions of $x$.\footnote{Because $v$
and $v'$ belong to $\cC_{\alpha>1/2}$, $\mu_\pm$ and $\nu$ are
bounded functions for $x>a$ and $a_0$ sufficiently large.} By virtue
of (\ref{H-def-F}) and (\ref{bH-def}), this implies that the entries
of $\sH_u(x)$ and consequently $\bH(x)$ are products of $u(x)$ and
certain bounded functions of $x$. Therefore, there is a positive
real number $\beta$ such that the entries $H_{ij}(x)$ of $\bH(x)$
satisfy
    \be
    |H_{ij}(x)|\leq \beta k^{-1}|u(x)|.
    \label{bound-1}
    \ee

Now, let $\bU^{(\ell)}$ denote the $\ell$-th term in the Dyson series expansion (\ref{dyson}) of $\bU(\infty,a)$, i.e.,
    \be
    \bU^{(\ell)}:=\int_{a}^\infty dx_\ell \int_{a}^{x_\ell}dx_{\ell-1}\cdots
    \int_{a}^{x_{2}}dx_{1}[\bH(x_\ell)\bH(x_{\ell-1})\cdots\bH(x_1)],
    \label{Uell=}
    \ee
and $U^{(\ell)}_{ij}$ label its entries. Then, in view of (\ref{bound-2}), (\ref{bound-1}), and the fact that the entries of the matrix $\bH(x_\ell)\bH(x_{\ell-1})\cdots\bH(x_1)$ have the form $\sum_{k_1=1}^2\sum_{k_2=1}^2\cdots\sum_{k_\ell=1}^2H_{ik_1}H_{k_1k_2}\cdots H_{k_\ell j}$, we infer
    \bea
    |U^{(\ell)}_{ij}|&\leq&\left(\frac{2\beta}{k}\right)^{\!\!\ell}
    \int_{a}^\infty dx_\ell \int_{a}^{x_\ell}dx_{\ell-1}\cdots
    \int_{a}^{x_{2}}dx_{1}|u(x_\ell)u(x_{\ell-1})\cdots u(x_1)|\nn\\
    &\leq&\frac{1}{\ell!}\left(\frac{2\beta}{k}\right)^{\!\!\ell}
    \int_{a}^\infty dx_\ell \int_{a}^\infty dx_{\ell-1}\cdots
    \int_{a}^\infty dx_{1}|u(x_\ell)u(x_{\ell-1})\cdots u(x_1)|\nn\\
    &\leq&\frac{1}{\ell !}\left[\frac{2\beta\gamma}{\delta (ka)^\delta}\right]^\ell.
    \eea
This shows that the error associated with the approximation,
    \be
    \bU(\infty,a)\approx\bI+\sum_{\ell=1}^n \bU^{(\ell)},
    \label{approx-n}
    \ee
is proportional to $(ka)^{-n\delta}$. Hence, we can reduce it by adopting
larger values of $a$.

Let us examine the first-order approximation. Substituting (\ref{Uw-GG}) in (\ref{bH-def}) and making use of (\ref{Uell=}) and (\ref{approx-n}) with $n=1$, we find
    \bea
    \bH(x)&=&\brv\bcG_w(a)\,\bcH(x)\brv\bcG_w(a)^{-1},
    \label{bH=2}\\[6pt]
    \bU(\infty,a)&\approx&\bI+\brv\bcG_w(a)\int_{a}^\infty \bcH(x)dx\:\brv\bcG_w(a)^{-1},
    \label{U1=}
    \eea
where
    \begin{align}
    &\bcH(x):=\frac{u(x)}{2k}\,\bG_u(x)^{-1}\bcK\,\bG_u(x),
    \nn\\
    &\bG_u(x):=e^{-i\varsigma_u(x)\bsigma_3}\bG(x) e^{iS(x)\bsigma_3}.
    \nn
    \end{align}
Now, consider setting $n=1$ in (\ref{approx-n}). This amounts to ignoring quadratic and higher order terms in powers of $(ak)^{-\delta}$. With the help of (\ref{bound-2}), we observe that
    \be
    |\varsigma_u(x)|\leq\frac{1}{2k}\int_a^x |u(s)|ds\leq \frac{1}{2k}\int_a^\infty |u(s)|ds
    \leq\frac{\gamma}{\delta (ak)^\delta}.\nn
    \ee
Therefore, in computing the right-hand side of (\ref{U1=}), which
involves $u(x)e^{\pm i\varsigma_u(x)\bsigma_3}$, we can approximate
these terms by $u(x)\bI$. In view of this observation and the
identities, $\bcK\,\bsigma_1=-\bsigma_1\bcK=\bsigma_1$ and
$\bcK^2=\bzero$, (\ref{U1=}) gives
    \be
    \bU(\infty,a) \approx \bI+\brv\bcG_w(a)\,\bcU\,\brv\bcG_w(a)^{-1},
    \label{approx-1}
    \ee
where
    \bea
    \bcU&:=&\frac{1}{2k}\int_a^\infty \frac{u(x)}{\tau(x)}\,e^{-iS(x)\bsigma_3}\bcK\,
    e^{iS(x)\bsigma_3} dx=\left[\begin{array}{cc}
    \cU_0 & \cU_-\\
    -\cU_+ & -\cU_0
    \end{array}\right],
    \label{bcU1=}\\
    \cU_0&:=&\frac{1}{2k}\int_{a}^\infty \frac{u(x)}{\tau(x)}\, dx,\quad\quad\quad
    \cU_\pm:=\frac{1}{2k}\int_{a}^\infty \frac{u(x)\, e^{\pm 2i S(x)}}{\tau(x)}\, dx.
    \eea
Because $u$ and $\IM(v)$ belong to $\cC_{\alpha>1}$, and
$S(x)=kx+\varsigma(x)$, the improper integrals yielding $\cU_0$ and
$\cU_\pm$ converge.

Substituting (\ref{approx-1}) in (\ref{M=MU}), we obtain the following approximate expression for the transfer matrix of $v_{+}$.
    \be
    \brv\bM_{{+}}\approx (\bI+\bcU)\brv\bcG_w(a)^{-1}.
    \label{M=MU-approx}
    \ee
In view of (\ref{Mw=}) and the fact that
$e^{-i\varsigma_u(\infty)\bsigma_3}-\bI$ contributes as a
first-order term in our perturbation scheme, we can also express
(\ref{M=MU-approx}) in the form,
    $\brv\bM_{{+}}\approx \brv\bM_{{+}}^{(0)}+\brv\bM_{{+}}^{(1)}$, where
    \begin{align}
    &\brv\bM_{+}^{(0)}:=\brv\bM_{w},
    &&    \brv\bM_{+}^{(1)}:=\left[\,\bcU -i \varsigma_u(\infty)\,\bsigma_3\right]\brv \bM_{w}.\nn
    \end{align}
Clearly, $\brv\bM_{{+}}\approx \brv\bM_{{+}}^{(0)}$ gives the zeroth-order approximation corresponding to $\bU(\infty,a)\approx\bI$.

\begin{figure}
    \begin{center}
    \includegraphics[scale=.42]{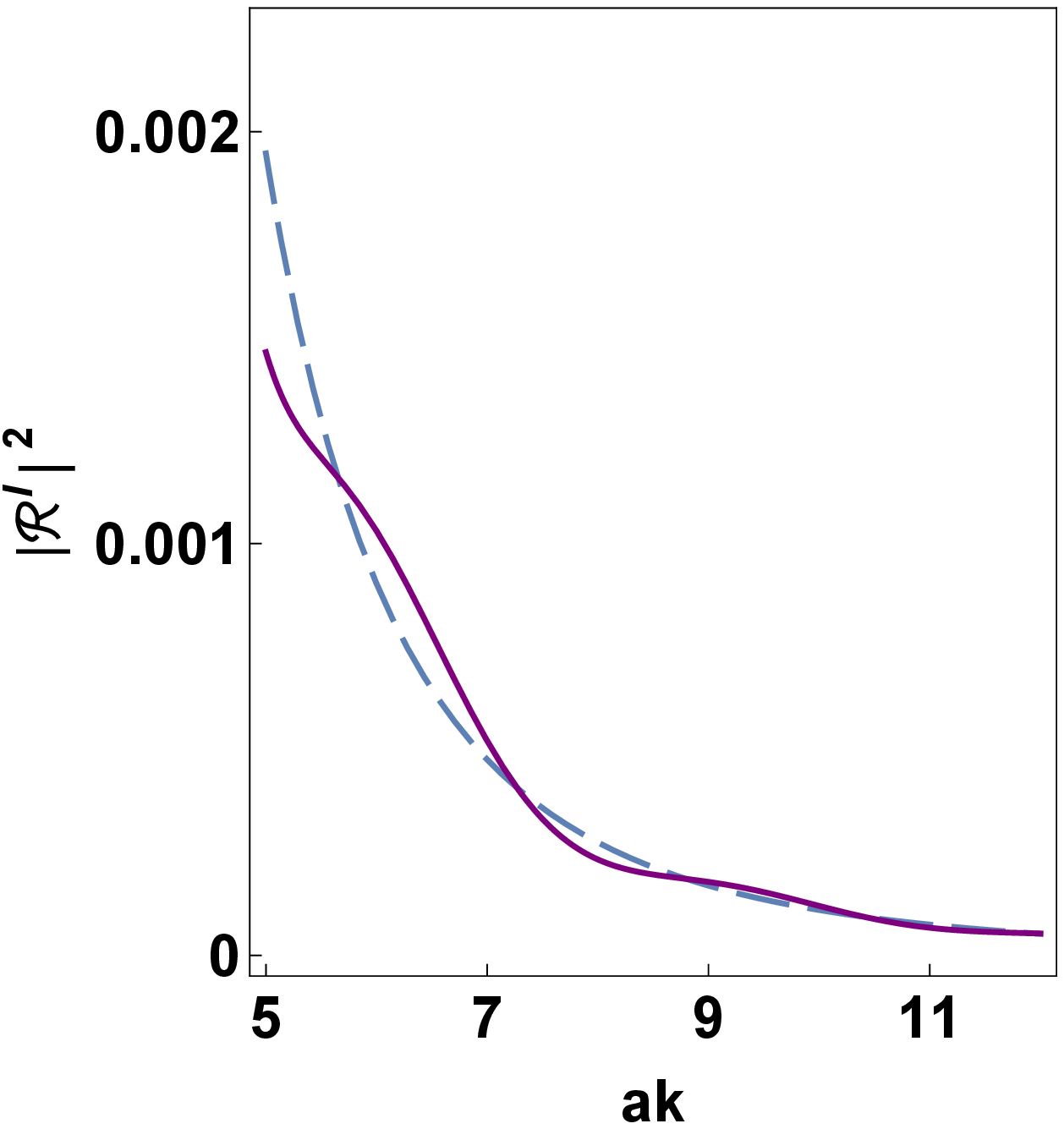}~~~~
    \includegraphics[scale=.42]{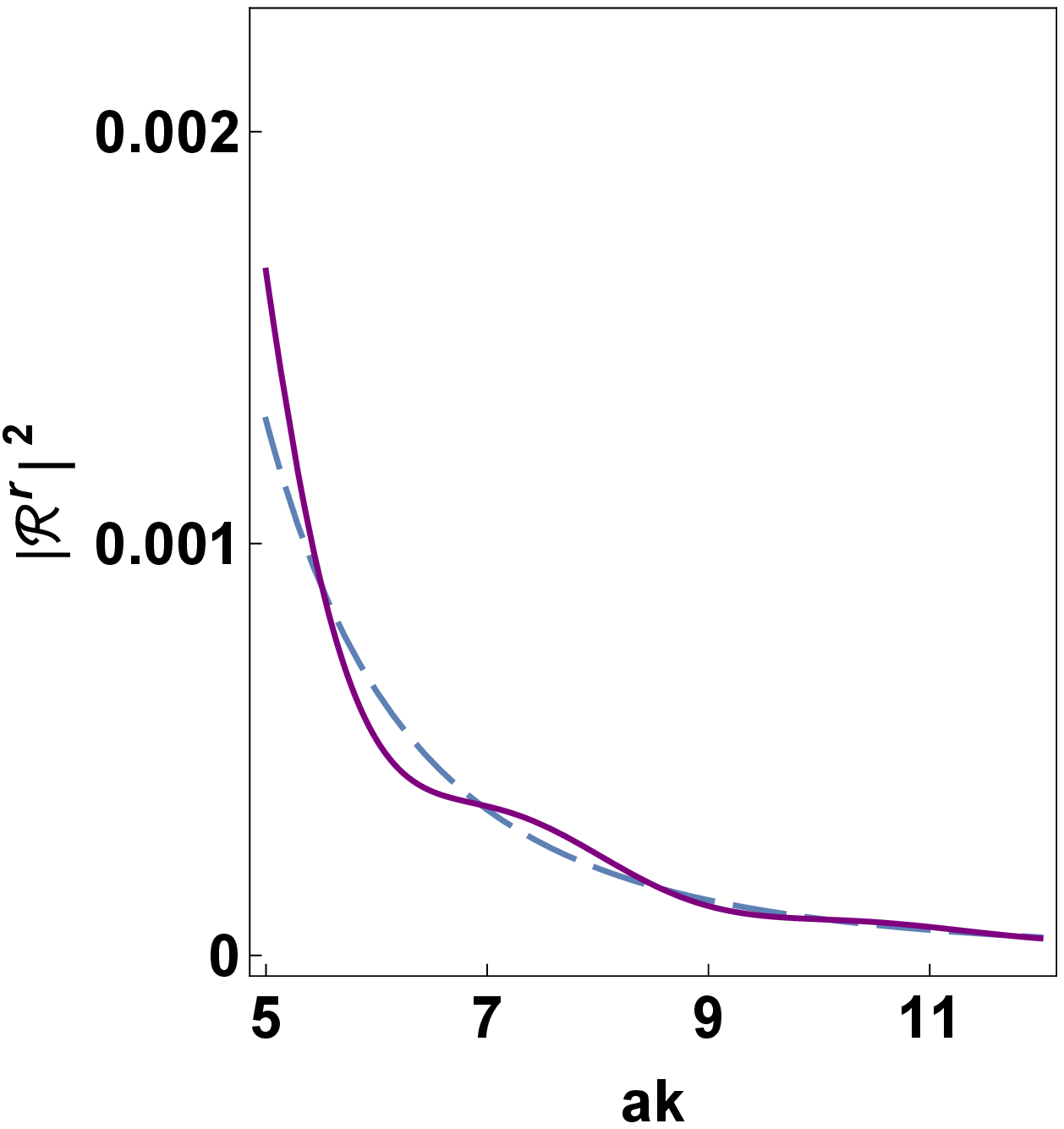}~~~~
    \includegraphics[scale=.41]{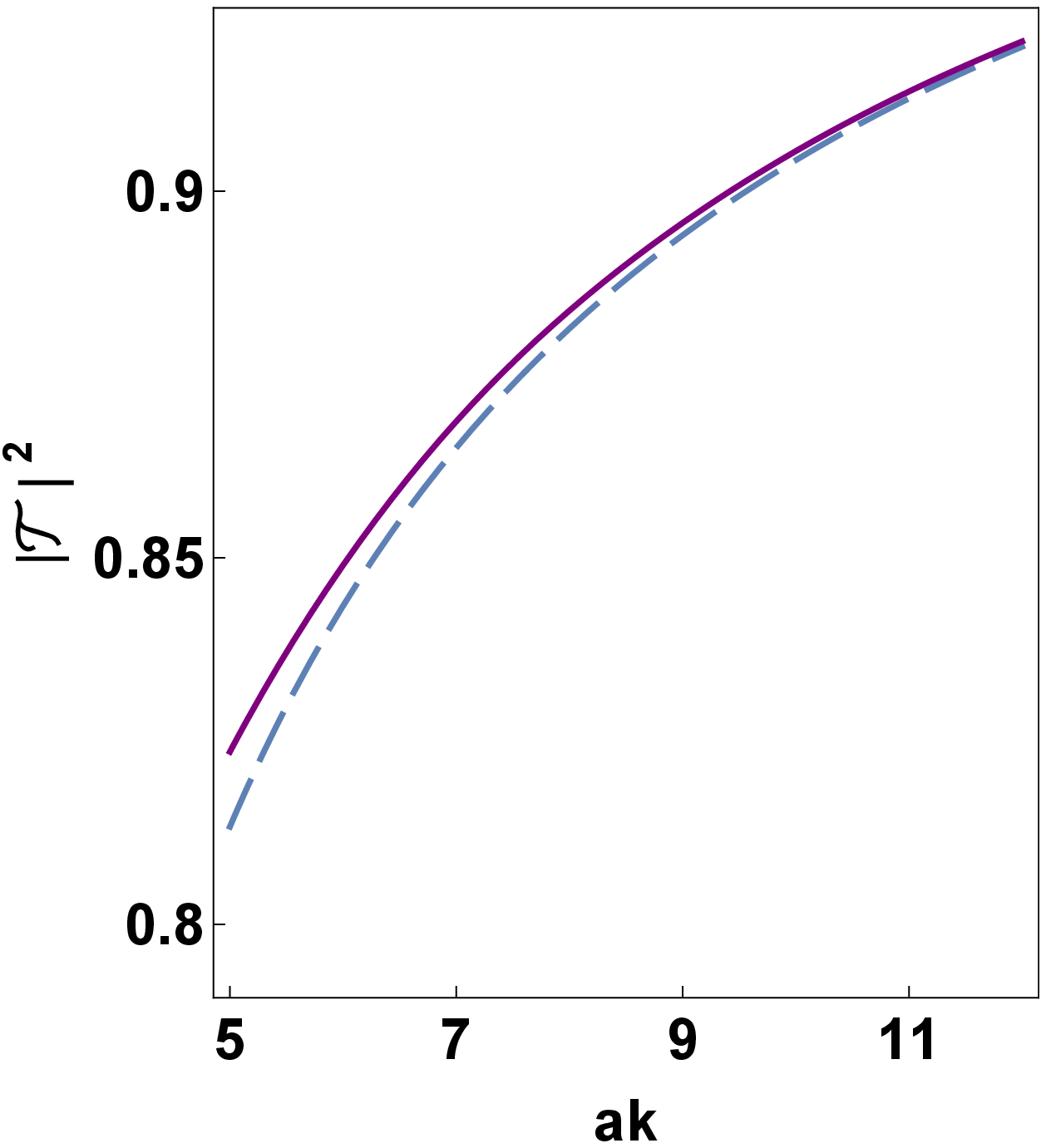}
    \caption{Plots of $|\cR^l|^2$, $|\cR^{r}|^2$, and $|\cT|^2$ as functions of $ak$ for the potential (\ref{v-toy1}) with $g=-a^{-1}$ and $\fz=5-i$. The dashed (dark blue)  and solid (purple) curves respectively correspond to the results of the zeroth- and first-order perturbative calculations. They converge for larger values of $ak$.}
    \label{fig1}
    \end{center}
    \end{figure}

As an example consider the potential,
    \be
    v(x)=\frac{g}{x}+\frac{\fz}{x^2},
    \label{potential=}
    \ee
where $g$ and $\fz$ are respectively real and complex coupling constants. It clearly satisfies (\ref{condi-complex}). Therefore, whenever $a|g|+|\fz|<2(ak)^2$ and $ak\gg 1$,  we can use the above perturbation scheme to determine the transfer matrix $\brv\bM_+$ and the reflection and transmission amplitudes, $\cR^{l/r}$ and $\cT$, of the potential:
    \be
    v_+(x)=\left\{\begin{array}{ccc}
    \displaystyle\frac{g}{x}+\frac{\fz}{x^2} & {\rm for} & x\geq a,\\[6pt]
    0 & {\rm for} & x<a.
    \end{array}\right.
    \label{v-toy1}
    \ee
Fig.~\ref{fig1} shows the graphs of the reflection and transmission coefficients, $|\cR^{l/r}|^2$ and $|\cT|^2$, of this potential for $g=-1/a$, $\fz=5-i$, and $ak\geq 5$. The dashed and solid curves correspond to the results of the zeroth- and first-order perturbative calculations, respectively. As expected, their difference diminishes as $ak$ grows.

Let us recall the decomposition (\ref{decompose-v}) of the potential $v$. We can use the above perturbative approximation scheme to compute the transfer matrix $\brv\bM_-$ of the potential $v_-$ and use the composition rule for transfer matrices to determine the transfer matrix $\brv\bM$ of the potential $v$. A desirable aspect of this  approximation scheme is that we can improve its accuracy not only by including higher order terms in the perturbative expansion of $\brv\bM_\pm$, but also by choosing larger values of $a$ which would reduce the perturbation parameter $(ak)^{-1}$. Indeed for every value of $k$, we can adjust $a$ so that $ak$ attains such a large value that even the zeroth-order approximation is valid. This marks an important distinction between our scheme and the standard WKB approximation which is generically valid for high energies. The price one pays for taking large values of $a$ is to increase the size of the support of the potential $v_0$. This can in principle complicate the computation of its transfer matrix. But, $v_0$ has a finite range, and there are well-known numerical methods for a direct or indirect determination of its transfer matrix.

\section{Generating exactly solvable long-range potentials}
\label{S6}

A basic property of second order linear homogeneous ordinary differential equations is that given a nonzero solution of this equation we can obtain a second linearly independent solution \cite{ODE}. In this section, we use this property to generate an exactly solvable long-range potential of the form (\ref{v-toy1}). Finding an exact expression for the general solution of the Schr\"odinger equation (\ref{sch-eq}) for this potential is equivalent to the exact solution of (\ref{sch-eq}) for the potential $v$ in the interval $[a,\infty)$.

Let $\phi_+:[a,\infty)\to\C$ be a function of the form,
    \be
    \phi_+(x):=e^{\xi(x)}e^{iS_{v_+}\!(x)},
    \label{psi-1-def}
    \ee
where $\xi:[a,\infty)\to\C$ is an axillary function,
$S_{v_+}(x):=kx+\varsigma_+(x)$, and
    \be
    \varsigma_+(x):=-\frac{1}{2k}\int_0^x v_+(s) ds=-\frac{1}{2k}\int_a^x v(s) ds.
    \label{chi-p-def}
    \ee
Demanding $\phi_+$ to solve the Schr\"odinger equation,
    \be
    -\psi''(x)+[v(x)+\fq(x)]\psi(x)=k^2\psi(x),
    \label{sch-eq-nu}
    \ee
in the half-line $[a,\infty)$, we find
    \be
    \fq=\xi''+\xi'^2+2ik\left(1-\frac{v}{2k^2}\right)\xi'-\frac{v^2}{4k^2}-\frac{iv'}{2k}.
    \label{nu=}
    \ee

$\phi_+$ is a solution of the Schr\"odinger equation (\ref{sch-eq}) for the potential $v$ provided that we select $\xi$ in such a way that $\fq=0$. For a potential of the form (\ref{potential=}), we can satisfy this equation using the ansatz
    \be
    \xi(x)=\frac{\fc}{x},
    \label{mu=}
    \ee
where $\fc$ is a constant. Substituting  (\ref{potential=}) and (\ref{mu=}) in (\ref{nu=}) and demanding its right-hand side to vanish, we obtain
    \begin{align}
    &\fc=\fc_\star:=\frac{g (2k+ig)}{8 k^3},
    \label{fa=}\\
    &\fz=\fz_\star:=2ik\,\fc=\frac{g (-g + 2i k)}{4 k^2}.
    \label{fz=}
    \end{align}
These in turn imply
    \bea
    S_{v_+}(x)&=&kx-\frac{1}{2k}\left[g\ln\left(\frac{x}{a}\right)+
    \frac{\fz(x-a)}{ax}\right],\\
    \phi_+(x)&=&e^{\fc_\star/a}\exp\left\{i[kx-(g/2k)\ln(x/a)]\right\}.
    \label{psi-p=}
     \eea
The latter is an exact solution of the Schr\"odinger equation
(\ref{sch-eq}) for the potential (\ref{potential=}) in $[a,\infty)$
provided that (\ref{fa=}) holds. According to
(\ref{psi-1-def}) and (\ref{mu=}), $\phi_+$ has the appealing property:
    \be
    \phi_+(x)\to e^{iS_{v_+}\!(x)}~~~{\rm for}~~~x\to\infty.
    \label{phi-p-asym}
    \ee
In Appendix we construct another solution, $\phi_-$, of the same Schr\"odinger equation that satisfies
    \be
    \phi_-(x)\to e^{-iS_{v_+}\!(x)}~~~{\rm for}~~~x\to\infty.
    \label{phi-m-asym}
    \ee
It is given by
    \be
    \phi_-(x):=\frac{1}{\phi_+(x)}+\frac{ig}{k}\,\phi_+(x)\!\int_x^\infty\!\frac{ds}{s\,\phi_+(s)^2}.
    \label{psi-m=}
    \ee

Now, consider the potential (\ref{v-toy1}) with $\fz$
given by (\ref{fz=}), i.e.,
    \be
    v_+(x)=\left\{\begin{array}{ccc}
    \displaystyle\frac{g}{x}+\frac{\fz_\star}{x^2} & {\rm for} & x\geq a,\\[6pt]
    0 & {\rm for} & x<a.
    \end{array}\right.
    \label{v-star}
    \ee
The above analysis shows that the
corresponding Schr\"odinger equation (\ref{sch-eq}) admits a pair of
linearly independent solutions of the form (\ref{psi-pm=}) with
$f_\pm$ replaced with $\phi_\pm$. We can follow the approach of
Sec.~4 to express the transfer matrix $\brv\bM$ for this potential
in the form,
    \be
    \brv\bM_+=\left[\begin{array}{cc}
    \fb_- & -\fa_-\\
    -\fb_+ & \fa_+\end{array}\right],
    \ee
where $\fa_\pm$ and $\fb_\pm$ are given by (\ref{fa-def}) and
(\ref{fb-def}) with $f_\pm$ replaced with $\phi_\pm$. In view of
(\ref{fz=}), (\ref{psi-p=}), and (\ref{psi-m=}), this gives
    \begin{align}
    &\fa_-=\left[-\widehat g\, e^{-2iak}+2i\widehat g\left(1-\widehat g\right)\cI_0\right]e^{-\fc/a},
    &&\fa_+=\left(1-\widehat g\right)e^{\fc/a},
    \label{fas=2}\\
    &\fb_-= \left(1+\widehat g+2i{\widehat g}^{\,2}\, e^{2iak}\cI_0\right)e^{-\fc/a},
    &&\fb_+=\widehat g\, e^{2iak} e^{\fc/a},
    \label{fbs-2}
    \end{align}
where $\widehat g:=g/4ak^2$,
    \bea
    \cI_0&:=&2ak\,e^{2\fc/a}\!\!\int_a^\infty\!\frac{dx}{x\,\phi_+(x)^2}
    =(2ak)^{1-ig/k}\int_{2ak}^\infty t^{-1+ig/k} e^{-it}dt\nn\\
    &=&e^{\pi g/2k}(2ak)^{1-ig/k}\,\Gamma(ig/k,2iak),
    \label{I-zero=}
    \eea
and $\Gamma(\cdot,\cdot)$ stands for the incomplete Gamma function
\cite{Gradshteyn-Rezhnik}.

\begin{figure}
    \begin{center}
    \includegraphics[scale=.43]{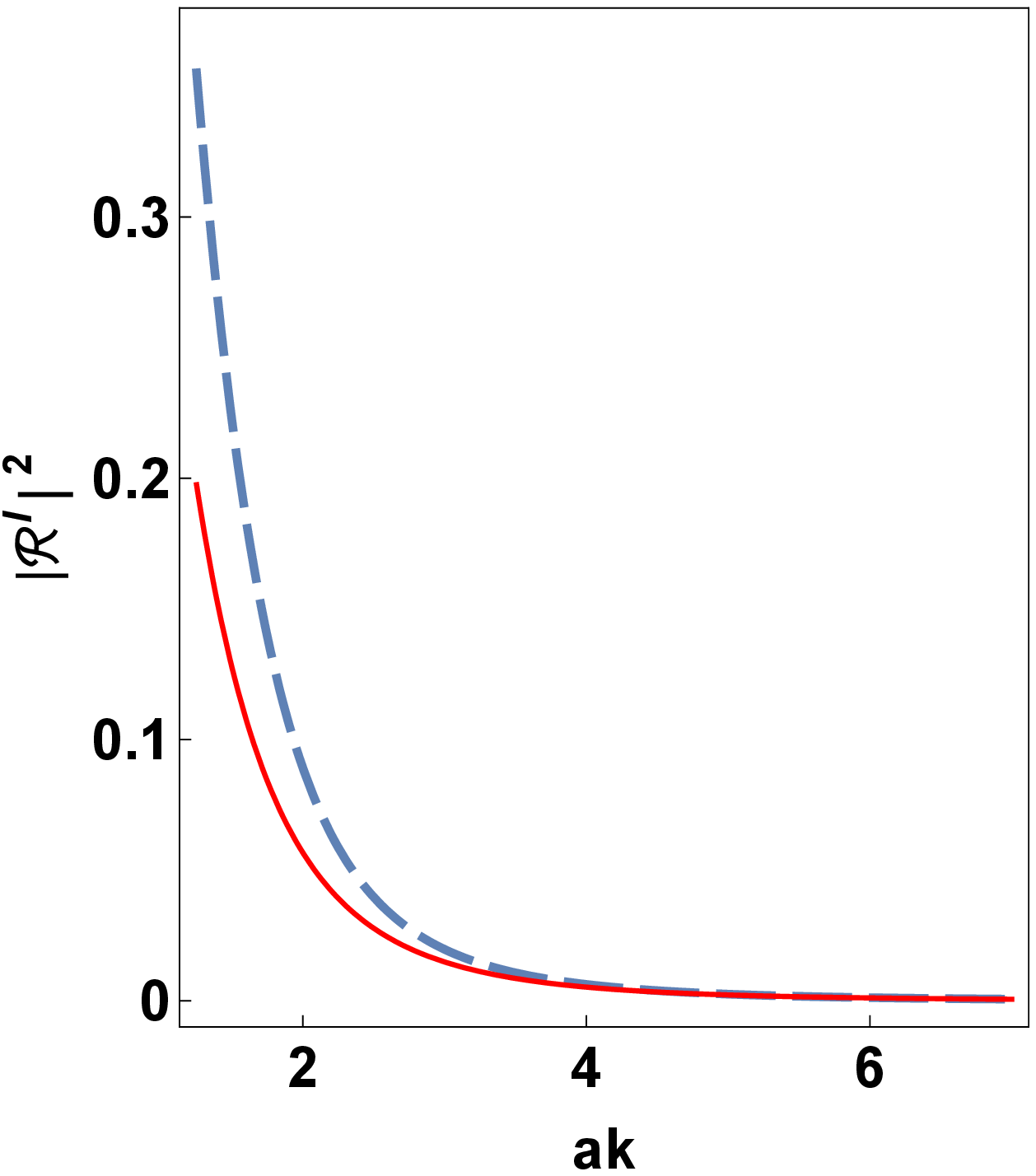}~~~~
    \includegraphics[scale=.43]{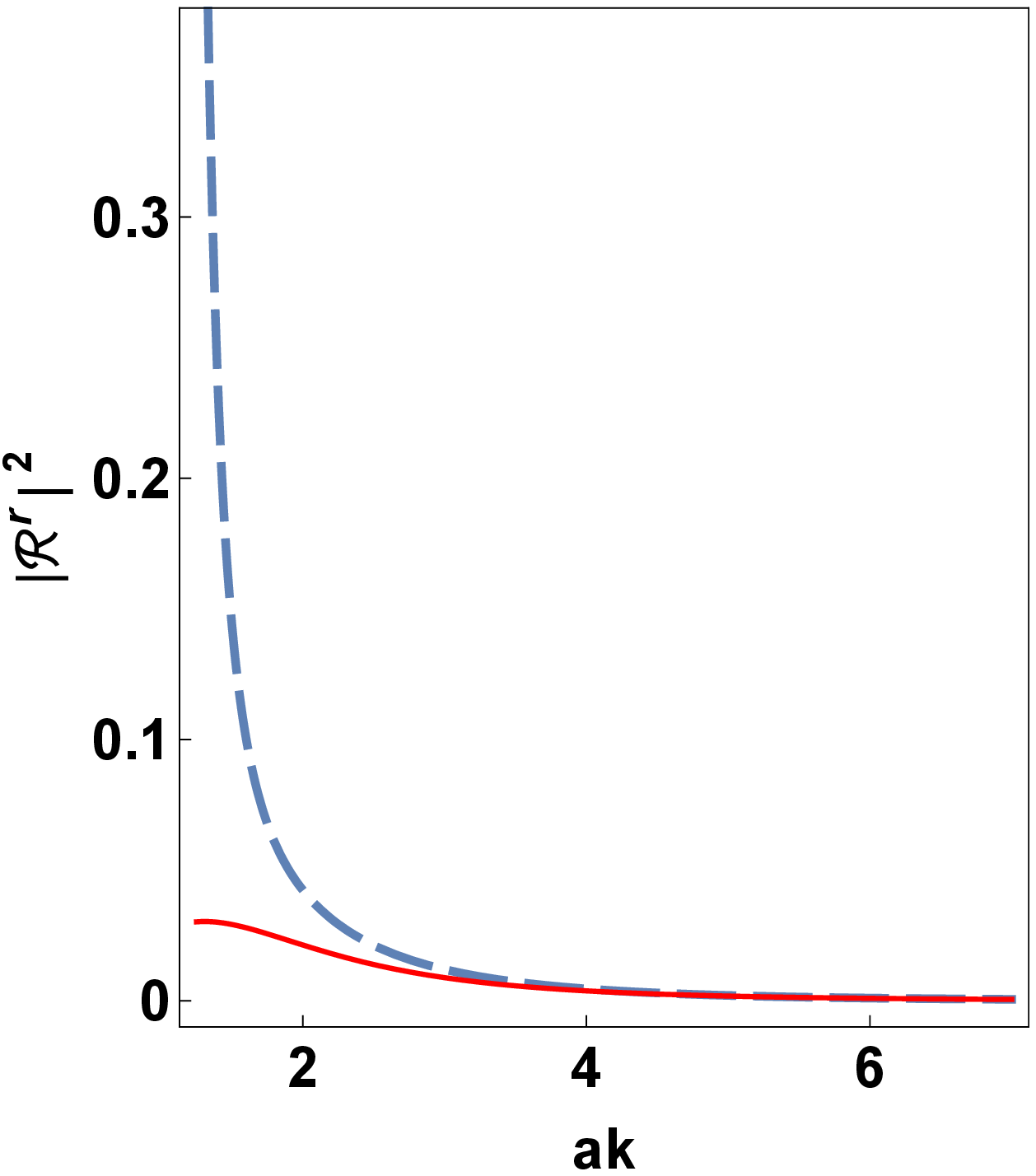}~~~~
    \includegraphics[scale=.42]{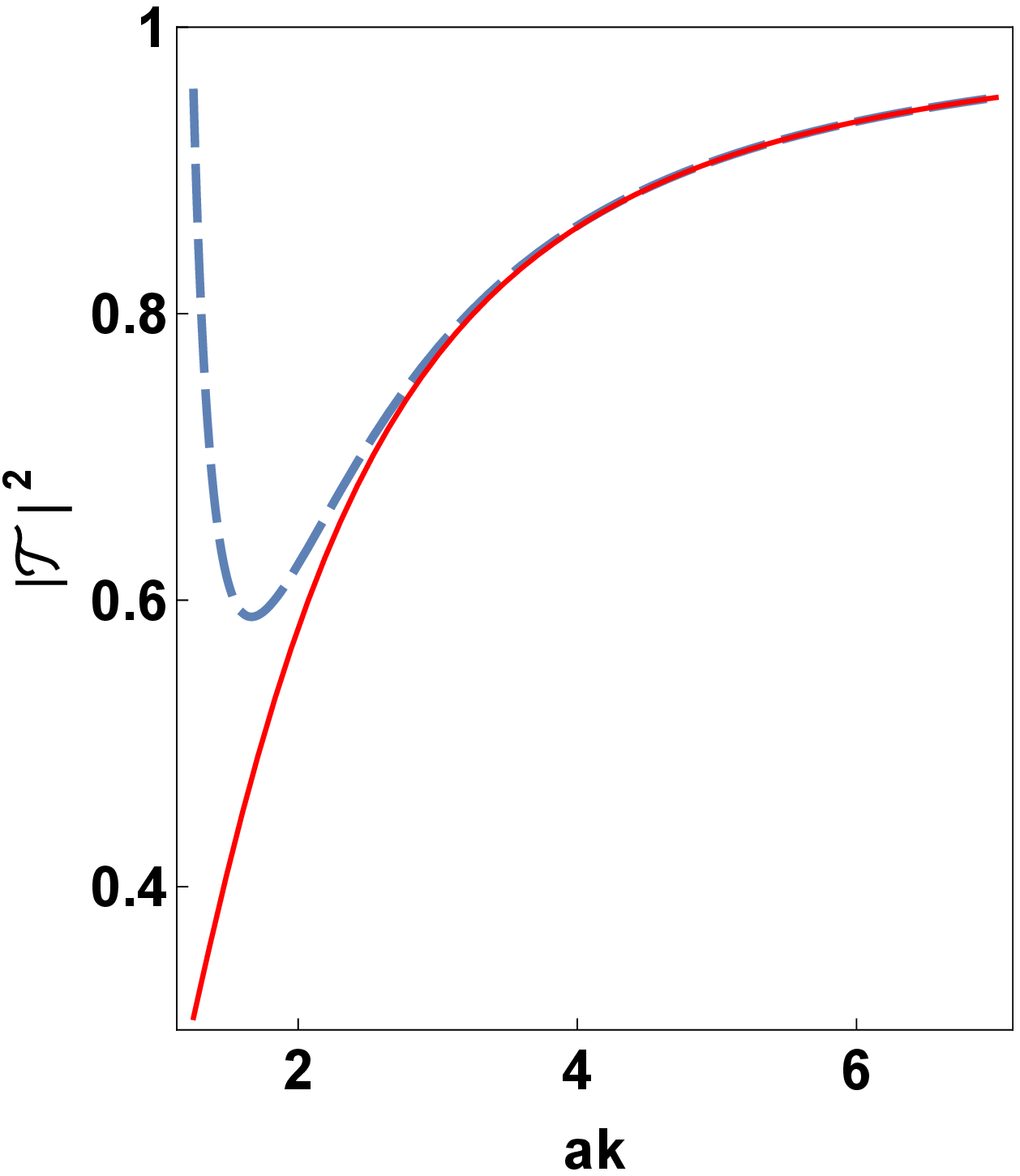}
    \caption{Plots of $|\cR^l|^2$, $|\cR^{r}|^2$, and $|\cT|^2$ as functions of $ak$ for the potential (\ref{v-toy1}) with $g=-5a^{-1}$ and $\fz$ given by (\ref{fz=}). The dashed (dark blue) and solid (red) curves respectively correspond to the results of the zeroth-order perturbative calculations of Sec.~\ref{S5} and the exact results given by Eqs.~(\ref{exact-RL}), (\ref{exact-RR}), and (\ref{exact-T}).}
    \label{fig2}
    \end{center}
    \end{figure}

Having obtained the transfer matrix $\brv\bM_+$, we use (\ref{RT-brv-c2}) to calculate the reflection and transmission amplitudes, $\cR^{l/r}$ and $\cT$, of the potential (\ref{v-star}). Because for this potential $\vartheta_i^-=0$ and $\vartheta_i^+=-\IM(\fz)/2ak=-\widehat g$, this yields
    \bea
    \cR^l&=&\frac{\fb_+}{\fa_+}=\frac{\widehat g\, e^{2iak}}{1-\widehat g},
    \label{exact-RL}\\
    \cR^r&=&-\frac{e^{2\widehat g}\fa_-}{\fa_+}=
    e^{-4iak \widehat g^2}
    \left(\frac{\widehat g\, e^{-2iak}}{1-\widehat g}
    -2i\widehat g\,\cI_0\right),
    \label{exact-RR}\\
    \cT&=&\frac{e^{\widehat g}}{a_+}=\frac{e^{-2iak \widehat g^2}}{1-\widehat g}
    \label{exact-T}.
    \eea
Fig.~\ref{fig2} shows plots of the reflection and transmission
coefficients for the potential (\ref{v-star}) with $g=-5a^{-1}$. For
large values of $ak$ the approximate results obtained using the
zeroth-order perturbation schemes of Sec.~4 are in perfect agreement
with the exact results provided by (\ref{exact-RL}) --
(\ref{exact-T}).

\section{Summary and concluding remarks}
\label{S7}

Transfer matrices have been extensively used in dealing  with
scattering problems since the 1940's. Their applications were
however confined to the study of short-range potentials. In this
article, we extended their domain of application to a large class of
real and complex long-range potentials. This involved a
re-examination of the standard notion of the transfer matrix of a
short-range potential $\bM$, its identification with the S-matrix of
a certain effective two-level quantum system, its relationship with
the classical notion of the fundamental matrix of the theory of
linear ordinary differential equations, and most notably the
introduction of  a pair of transfer matrices $\brv\bM$ and $\bcM$
for lang-range potentials which shared the basic features of $\bM$.
In particular, they store the information about the scattering
features of the potential and possess the same composition property.

We have employed the composition property of $\brv\bM$ to reduce the
problem of dealing with the long-range potentials of our interest to
those supported in an interval of the form $[a,\infty)$ with $a>0$.
We have then introduced a decomposition of such potentials into the
sum of a short-range potential and an exactly solvable potential.
This has motivated us to develop a perturbative approximation scheme
for the computation of the transfer matrix $\brv\bM$ whose accuracy
can be improved by choosing larger values of $a$.

In order to demonstrate the utility of our approximation scheme, we
have introduced an exactly solvable long-range complex potential and
compared the outcome of the exact and approximate calculations of
its reflection and transmission coefficients. Our explicit
calculations reveal an almost perfect agreement between the exact
and approximate results for $ak\gg 1$.

\section*{Appendix: Construction of $\phi_-$}

Consider the solution $\phi_+$ of the Schr\"odinger equation
(\ref{sch-eq})  for the potential (\ref{potential=}) with $\fz$
given by (\ref{fz=}). We can use $\phi_+$ to express the general
solution of (\ref{sch-eq}) in $[a,\infty)$ as \cite{ODE},
    \be
    \psi=c_+ \phi_+ + c_-\psi_-,
    \label{gen-sol-app}
    \ee
where $c_\pm$ are constant coefficients,
    \be
    \psi_-(x):=\phi_+(x)\int_b^x \frac{ds}{\phi_+(s)^2}.
    \label{psi-p-def}
    \ee
and $b$ is a real number exceeding $a$. In view of (\ref{psi-1-def}),
    \bea
    \psi_-(x)&=&
    \phi_+(x)\int_b^x e^{-2[\xi(s)+i\varsigma_+(s)]}e^{-2iks}ds\nn\\
    &=&\phi_+(x)\left[\left.\frac{i}{2k\: \phi_+(s)^2}\right|^x_b
    +\frac{i}{k}\int_b^x \frac{\xi'(s)+i\varsigma'_+(s)}{\phi_+(s)^2}\,ds\right]\nn\\
    &=&\frac{i}{2k}\left[\frac{1}{\phi_+(x)}-\frac{\phi_+(x)}{\phi_+(b)^2}
    +\frac{ig}{k}\,\phi_+(x)\!\int_x^b\!\frac{ds}{s\,\phi_+(s)^2}
    \right],\nn
    \eea
where we have performed an integration by parts and employed (\ref{potential=}), (\ref{chi-p-def}), (\ref{mu=}), and (\ref{fz=}). Now, let $c_-=-2ik$ and $c_+=\phi_+(b)^{-2}$. Then for each $b>a$, (\ref{gen-sol-app}) produces the following solution of the Schr\"odinger equation (\ref{sch-eq}) for the potential (\ref{potential=}) in $[a,\infty)$.
    \be
    \psi_b(x):=\frac{1}{\phi_+(x)}+\frac{ig}{k}\,\phi_+(x)\!\int_x^b\!\frac{ds}{s\,\phi_+(s)^2}.\nn
    \ee

With the help of (\ref{psi-p=}) we can show that
$\int_x^b\!\frac{ds}{s\,\phi_+(s)^2}=\fh\,[\cI(2kb)-\cI(2kx)]$,
where $\fh:=e^{-2\fc/a}(2ak)^{-ig/k}$, $\cI(y):=\int_{1}^y
t^{-1+ig/k}e^{-i t}dt=e^{\pi g/2k}\Gamma(ig/k,i,iy)$, and
$\Gamma(\cdot,\cdot,\cdot)$ is the generalized incomplete Gamma
function. It is not difficult to see that for $y\geq 1$,
    \bea
    \left|\cI(y)\right|&=&\left| y^{-1+ig/k}e^{-iy}-e^{-i}+
    \left(1-\frac{ig}{k}\right)\int_1^y t^{-2+ig/k}e^{-it}dt\right| \nn\\
    &\leq & \frac{1}{y}+1+\left(1+\frac{g^2}{k^2}\right)
    \left|\int_1^y t^{-2+ig/k}e^{-it}dt\right|\nn\\
    &\leq&\frac{1}{y}+1+\left(1+\frac{g^2}{k^2}\right)\int_1^y \frac{dt}{t^{2}}
    =2+\frac{g^2}{k^2}\left(1-\frac{1}{y}\right).
    \label{bound-3}
    \eea
where we have used integration by parts in the first line and
benefitted from the fact that $g$ is real. According to
(\ref{bound-3}), $\cI(\infty):=\lim_{y\to\infty}\cI(y)$
exists\footnote{In fact, $\cI(\infty)=e^{\pi g/2k}\Gamma(ig/k,i)$
where $\Gamma(\cdot,\cdot)$ is the incomplete Gamma function.} and
the improper integral $\int_x^\infty\!\frac{ds}{s\,\phi_+(s)^2}$
converges. This in turn implies that the $\phi_-$ given by
(\ref{psi-m=}) is a well-defined function in the interval
$[a,\infty)$. It is easy to check that it solves the Schr\"odinger
equation (\ref{sch-eq}) for the potential (\ref{potential=}) with
$\fz$ given by (\ref{fz=}) and that it satisfies (\ref{phi-p-asym}).

\section*{Acknowledgements}
We thank Turkish Academy of Sciences (T\"UBA) for supporting FL's visit to Ko\c{c} University in 2019 during which this work was initiated. AM has been supported by T\"UBA's membership grant.

\ed

It has the following explicit form,
    \be
    \phi_+(x)=\left(\frac{x}{a}\right)^{-ig/2k}e^{ikx-\fz(x-a)/2ak x}.
    \label{psi-p=2}
    \ee